\documentclass[twocolumn,preprintnumbers,showpacs]{revtex4}
\usepackage{amsmath}
\usepackage{extarrows}
\usepackage{graphicx}
\usepackage{dcolumn}
\usepackage{bm}
\usepackage[all]{xy}
\usepackage{indentfirst}
\usepackage{multirow}
\usepackage{amssymb}
\usepackage{lipsum}

\begin{document}
\title{Unified quantum no-go theorems and transforming of quantum states in a restricted set}
 \author{Ming-Xing Luo$^{1,2}$, Hui-Ran Li$^{2}$, Hong Lai$^{3}$, Xiaojun Wang$^{4}$}

\affiliation{
$^1$ Information Security and National Computing Grid Laboratory, Southwest Jiaotong University, Chengdu 610031, China
\\
${}^{2}$ Department of Physics, University of Michigan, Ann Arbor, MI 48109, USA
\\
$^3$ School of Computer and Information Science, Southwest University, Chongqing 400715, China
\\
$^4$ School of Electronic Engineering, Dublin City University, Dublin 9, Ireland}

\begin{abstract}
The linear superposition principle in quantum mechanics is essential for several no-go theorems such as the no-cloning theorem, the no-deleting theorem and the no-superposing theorem. It remains an open problem of finding general forbidden principles to unify these results. In this paper, we investigate general quantum transformations forbidden or permitted by the superposition principle for various goals. First, we prove a no-encoding theorem that forbids linearly superposing of an unknown pure state and a fixed state in  Hilbert space of a finite dimension. Two general forms include the no-cloning theorem, the no-deleting theorem, and the no-superposing theorem as special cases. Second, we provide a unified scheme for presenting perfect and imperfect quantum tasks (cloning and deleting) in a one-shot manner. This scheme may lead to fruitful results that are completely characterized with the linear independence of the input pure states. The generalized upper bounds for the success probability are proved. Third, we generalize a recent superposing of unknown states with fixed overlaps when multiple copies of the input states are available.

\end{abstract}


\renewcommand{\theequation}{\arabic{equation}}
\newtheorem{lm}{Lemma}
\newtheorem{thm}{Theorem}
\newtheorem{pro}{Proposition}
\newtheorem{cor}{Corollary}
\newtheorem{df}{Definition}
\newtheorem{exmp}{Example}
\newtheorem{rmk}{Remark}[section]

\maketitle

\section{Introduction}

In quantum mechanics, the no-cloning theorem as a well-known fact asserts that an arbitrary unknown pure state \cite{1} or noncommuting mixed state \cite{Barn} cannot be perfectly created using a universal quantum transformation. This non-trivial exception is derived from the superposition principle of quantum states and may be implied by the no-communication theorem which forbids the classical information transmission using only quantum entanglement. Although the no-cloning theorem prevents several important extensions of classical results such as classical error corrections \cite{shor,stean}, it plays a key role in various quantum applications, especially in quantum secure communication. The no-cloning theorem plays a key role in certifying the unconditional security of quantum cryptography \cite{2,Shor1,Lo,CBK,CS}. A universal cloning machine will violate the no-teleportation theorem that forbids converting a quantum state into classical bits. Unitary cloning also implies the possibility of deleting quantum information which is impossible. In fact, with the quantum superposition principle, unknown quantum states of a finite dimension cannot be unitarily transformed into a fixed state, i.e., the quantum no-deleting theorem \cite{3}. In general, the no-cloning theorem has no relationship to the no-deleting theorem because the post-selection permitted in the cloning task is forbidden in the deleting task.

The key of the no-cloning theorem and the no-deleting theorem is the superposition principle of quantum states \cite{4}. It is also essential for nontrivial quantum applications including Shor's factoring algorithm \cite{5}, Grover's search algorithm \cite{6}, quantum cryptography \cite{7}, quantum metrology \cite{8}, and Boson sampling \cite{9}. However, this principle is unsuitable for the task of superposing unknown states. A recent theorem  \cite{10,11} shows that it is impossible to create a superposition state of unknown states using a universal quantum transformation \cite{10}. This no-go theorem is different from the no-cloning theorem \cite{1}, the no-deleting theorem \cite{3} or the special no-superposing theorem that forbids superposing an unknown state and its orthogonal complement \cite{12}. It remains a challenge to explore intrinsic connections among these no-go theorems.

In comparison to recovering perfect resultant, Bu\v{z}ek and Hillery provided a universal quantum transformation for imperfect cloning \cite{18}. The partial trace on each copy of their resultant is a mixed state with the optimal fidelity \cite{Gis,Bru,Bru2,Kang}. Their scheme had triggered lots of investigations on quantum cloning tasks that include $N\to M$ cloning of qubits \cite{Gis}, $d$-dimensional universal cloning \cite{BH,Wer,Zav,Wang}, asymmetric cloning \cite{NG,Cerf,Buzek,Ibl,Cwi}, mixed-state cloning \cite{Barn,Dari,Chen,Dang}, experimental cloning \cite{De,Lin,Du,Zhao,Naga,Haw,WD,Zhu}, and relative discussions and applications \cite{19,Fan}. The cloning fidelities of these imperfect schemes may be improved to unit if input states are chosen from a finite set of the whole space. In particular, the linear independence of input pure states has been proved to be a sufficient and necessary condition for perfect cloning with a unitary and post-selection process \cite{13,14}. The optimal success probability is the same as that of state discriminations \cite{CB,7b,CJW,MW}. The linear independence has also been used for the task of superposing multiple copies of input states \cite{15}. Similar results hold for perfect deleting. These perfect schemes may be unified in a one-shot manner using multiple copies of input states \cite{16}.

Recently, a constructive scheme \cite{10} shows that a universal superposing transformation exists for unknown states chosen from two sets where all the states of each set have a fixed overlap with a known state. The scheme shows a new insight of the linear independence, i.e., a superposing scheme exists for a collection (generally uncountable) of finite sets consisting of linearly independent states. This is a great improvement for the quantum tasks defined on a finite set of quantum states. It should be interesting to investigate whether or not similar generalizations hold for other tasks of cloning or deleting \cite{13,14,15,16}.

In this paper, our motivation is to generalize the no-cloning theorem \cite{1}, no-deleting theorem \cite{3} and no-superposing theorem \cite{10}. Although they are different in general and have no explicit relationships among them, we hope to present a unified form which may provide new insights of these no-go theorems. Moreover, new constructions will be proposed to generate different desired states using the linear independence of input states that belong to restricted sets. The main contributions of this paper are listed as follows:
\begin{itemize}
\item[(1)]We present new no-go theorems. The first one is the no-encoding theorem that forbids linearly superposing of an unknown pure state and a fixed state (known or unknown) in the same space of a finite dimension from a universal quantum transformation. Its generalizations include the no-cloning theorem, the no-deleting theorem and the no-superposing theorem as special cases.
\item[(2)]We present several schemes to unify the perfect and imperfect tasks of cloning and deleting for pure states in a finite set. All the schemes are completely characterized by the linear independence of input states. Generalized upper bounds are proved using the matrix norm and state metric, respectively.
\item[(3)]We propose a scheme to superpose pure states in a finite set. We also present some generalized superposing schemes when multiple copies of input states are available.
\end{itemize}

The rest of the paper is organized as follows. In Sec. II, we present new no-go theorems. The first one is the no-encoding theorem. Namely, an arbitrary unknown state and a fixed state in the same space of a finite dimension cannot be linearly superposed using a universal quantum transformation with a nontrivial probability. This theorem will be further extended for multiple copies of the pure states. In Sec. III, several quantum transformations will be presented to unify imperfect and perfect tasks of cloning and deleting with controllable errors. The generalized bounds of the success probability will be proved using the matrix norm and state metric, respectively. Sec. IV contributes new superposing schemes that create the linear superposition of unknown pure states chosen from a finite set, or a restricted subspace when multiple copies of the input states are available. The last section concludes this paper.

\section{Unified no-go theorems}

In this section, we explore general forms of the no-cloning theorem, the no-deleting theorem and the no-superposing theorem.

\subsection{No-encoding theorem}

In what follows we only consider pure states. In quantum mechanics, a pure state provides a probability distribution for the value of each observable. In the mathematical formulation, from Riesz representation theorem \cite{ries} each pure state corresponds to a ray in a Hilbert space while each observable quantity is associated with a self-adjoint (or Hermitian) and positive semi-definite operator. In this paper, we only consider the normalized pure states. Thus each pure state is uniquely denoted by a density matrix $\rho_\psi$ on Hilbert space $\mathbb{H}$ with the dimension ${\rm dim}(\mathbb{H})=n\geq 2$, where $|\psi\rangle$ is its vector representative up to a global phase factor  $e^{i\theta}$, i.e, $\rho_{\psi}=\rho_{e^{i\theta}\psi}$. Denote the vector
$|\psi\rangle \propto |\phi\rangle$ if $\rho_{\psi}=\rho_{\phi}$, i.e., $|\psi\rangle=|\phi\rangle$ up to a global phase.

Let ${\cal CP}(\mathbb{H}_1, \mathbb{H}_2)$ be the set of completely positive (CP) maps which transform pure states on Hilbert space $\mathbb{H}_1$ into pure states on Hilbert space $\mathbb{H}_2$. All the unitary maps, which are denoted as $SU(\mathbb{H}_1)$, are CP maps that preserve the inner product of the Hilbert space. In particular, from Stinespring's Theorem \cite{Stin}, each ${\cal F}\in {\cal CP}(\mathbb{H}_1, \mathbb{H}_2)$ may be realized by a unitary map ${\cal U} \in SU(\mathbb{H}_1 \otimes \mathbb{H}_A\otimes \mathbb{H}_P)$ and a projection map ${\cal P}\in {\cal CP} (\mathbb{H}_1\otimes\mathbb{H}_A\otimes \mathbb{H}_P, \mathbb{H}_2)$, where $\mathbb{H}_A$ is an ancillary space and $\mathbb{H}_P$ is another ancillary space (named as probe space for simplicity) used for defining projectors. For each map ${\cal F}: \mathbb{H}_1\to \mathbb{H}_2$ with Hilbert spaces $\mathbb{H}_1$ and $\mathbb{H}_2$, denote ${\cal F}: \rho_{\psi}\mapsto \rho_{\phi}$ for $\rho_{\psi}$ on $\mathbb{H}_1$ and $\rho_{\phi}$ on $\mathbb{H}_2$ as its detailed mapping of ${\cal F}(\rho_{\psi})=\rho_{\phi}$.

\begin{thm}
Let $\alpha, \beta$ be nonzero complex constants satisfying $|\alpha|^2+|\beta|^2=1$, and $\rho_{\phi}$ be a fixed pure state (known or unknown) on Hilbert space $\mathbb{H}$ with ${\rm dim}(\mathbb{H})\geq 2$. There does not exist CP map ${\cal F}\in {\cal CP}(\mathbb{H}^2, \mathbb{H})$ for all pure states $\rho_{\psi}$ on $\mathbb{H}$ such that
\begin{eqnarray}
{\cal F}(\rho_{\psi}\rho_{\phi})=\rho_{\varphi},
\label{eqn-47}
\end{eqnarray}
where the vector representative of $\rho_{\varphi}$ is given by $|\varphi\rangle\propto\sqrt{r}(\alpha|\psi\rangle+\beta|\phi\rangle)$ and $r$ is a normalization constant dependent of $\alpha, \beta, |\psi\rangle$ and $|\phi\rangle$.

\label{thm4}
\end{thm}

The result in Theorem 1 holds for a fixed pure state $\rho_{\phi}$ which may be known or unknown (randomly chosen from $\mathbb{H}$). This theorem shows that one cannot use a fixed pure state to linearly superpose (or linearly encode) with an unknown state, which may be randomly chosen from $\mathbb{H}$ according to the Haar measure. It is also a general no-superposing theorem that forbids the creation of the linear superposition of an unknown state and a fixed state in the same space. Theorem 1 is not a corollary of the non-superposing Theorem \cite{10} while it is conversely true.

{\it Proof of Theorem 1}.  Note that each quantum operation can be represented by a unitary operation and post-selection of projective measurements \cite{14}. Thus a CP map is represented by a unitary transformation and the post-selection of projective measurements on a joint system consisting of the input states and ancillary states. Assume that there exists a CP map ${\cal F}$ satisfying Eq.(1). From Stinespring's Theorem \cite{Stin}, there exist two pure states $\rho_{\Sigma}$ (including the pure state $\rho_{\phi}$ as its subsystem) and $\rho_{\Sigma'}$ on ancillary Hilbert space $\mathbb{H}_A$, one probe state $\rho_{P}$ on probe space $\mathbb{H}_P$, a unitary map ${\cal U}: \mathbb{H}\otimes \mathbb{H}_A\otimes \mathbb{H}_P\to \mathbb{H}\otimes \mathbb{H}_A\otimes \mathbb{H}_P$ and a projective map ${\cal P}: \mathbb{H}\otimes \mathbb{H}_A\otimes \mathbb{H}_P\to \mathbb{H}\otimes \mathbb{H}_A$ such that ${\cal P}\circ {\cal U}(\rho_{\psi}\rho_{\Sigma}\rho_{P})=\rho_{\varphi}\rho_{\Sigma'}$ holds for all the states $\rho_{\psi}$ on $\mathbb{H}$. In the following, we only need to prove the nonexistence of the unitary map ${\cal U}$.

Now, assume there exists a unitary map ${\cal U}\in {\cal CP}(\mathbb{H}\otimes \mathbb{H}_A\otimes \mathbb{H}_P, \mathbb{H}\otimes \mathbb{H}_A\otimes \mathbb{H}_P)$ for all the states $\rho_{\psi}$ on  $\mathbb{H}$ such that
\begin{eqnarray}
{\cal U}(\rho_{\psi}\rho_{\Sigma}\rho_{P_0})=\rho_{\Psi},
\label{eqn-48}
\end{eqnarray}
where the vector representative of $\rho_{\Psi}$ is given by $|\Psi\rangle\propto \sqrt{p_0}\sqrt{r}(\alpha|\psi\rangle+\beta|\phi\rangle)
|\Sigma'\rangle_A|P_1\rangle+\sqrt{p_1}|\tilde{\Phi}\rangle$ and $r$ is a normalization constant. In Eq.(\ref{eqn-48}), $|\Sigma'\rangle$ is the vector representative of ancillary state $\rho_{\Sigma'}$ that may be dependent of $\alpha, \beta$, $|\psi\rangle$ and $|\phi\rangle$.  $|P_0\rangle$ and $|P_1\rangle$ are the respective vector representative of orthogonal pure states $\rho_{P_0}$ and $\rho_{P_1}$ on probe space $\mathbb{H}_P$. $p_0$ is the success probability of recovering a superposition state with the vector representative $\sqrt{r}(\alpha|\psi\rangle+\beta|\phi\rangle)$ that is dependent on $\alpha, \beta, |\psi\rangle, |\phi\rangle$, and satisfies $p_0>0$ and $p_0+p_1=1$. $|\tilde{\Phi}\rangle$ is the vector representative of a general failure state $\rho_{\tilde{\Phi}}$, that will be annihilated by the projector $|P_1\rangle\langle P_1|$, i.e, ${\rm tr}(\rho_{\tilde{\Phi}}|P_1\rangle\langle P_1|)=0$.

Here, we only consider the special case of $\rho_{\Sigma'}=\rho_{0}$. The proof of the general state $\rho_{\Sigma'}$ is shown in Appendix A. In this case, Eq.(\ref{eqn-48}) may be rewritten into
\begin{eqnarray}
U|\psi\rangle|\Sigma\rangle|P_0\rangle\propto
\sqrt{p_0}|\chi\rangle |0\rangle|P_1\rangle+\sqrt{p_1}|\tilde{\Phi}\rangle,
\label{eqq-3}
\end{eqnarray}
where $U$ is the matrix representative of the unitary mapping ${\cal U}$ and the vector representative of $\rho_{\chi}$ is defined by $|\chi\rangle\propto\sqrt{r}(\alpha|\psi\rangle+\beta|\phi\rangle)$.

For simplicity, we only consider $\rho_{\phi}=\rho_0$. A similar result may be proved using a generally fixed state $\rho_{\phi}$, which will be discussed in the next subsection. Since Hilbert space of qubit states is a subspace of general Hilbert space $\mathbb{H}$ with ${\rm dim}(\mathbb{H})>2$, it is sufficient to prove the nonexistence of Eq.(\ref{eqq-3}) for the qubit space. For the basis states $\rho_{0}, \rho_{1}$ on $\mathbb{H}$ (qubit space) as input states, Eq.(\ref{eqq-3}) leads to
\begin{eqnarray}
&&U|0\rangle|\Sigma\rangle|P_0\rangle\propto \sqrt{p_0}|0\rangle|0\rangle|P_1\rangle+\sqrt{p_1}|\tilde{\Phi}_0\rangle,
\label{eqq-4}
\\
&&U|1\rangle|\Sigma\rangle|P_0\rangle\propto \sqrt{p'_0}|\chi'\rangle|0\rangle|P_1\rangle+\sqrt{p'_1}|\tilde{\Phi}_1\rangle,
\label{eqq-5}
\end{eqnarray}
where the vector representative $|\chi'\rangle$ is defined by $|\chi'\rangle\propto \alpha|1\rangle+\beta|0\rangle$. In these equations, $p_0$ and $p'_0$ are the respective success probability of transforming pure state $\rho_{0}$ and $\rho_{1}$. $|\tilde{\Phi}_0\rangle$ and $|\tilde{\Phi}_1\rangle$ are the respective vector representative of the failure states $\rho_{\tilde{\Phi}_0}$ and $\rho_{\tilde{\Phi}_1}$ that should be annihilated by the projector $|P_1\rangle\langle P_1|$, i.e, ${\rm tr}(\rho_{\tilde{\Phi}_0}|P_1\rangle\langle P_1|)={\rm tr}(\rho_{\tilde{\Phi}_1}|P_1\rangle\langle P_1|)=0$.

For any pure state $\rho_h$ with the vector representative $|h\rangle\propto a|0\rangle+b|1\rangle$ with $|a|^2+|b|^2=1$, from Eq.(\ref{eqq-3}), we can obtain
\begin{eqnarray}
U|h\rangle|\Sigma\rangle|P_0\rangle\propto \sqrt{p_0''}|\varphi_1\rangle|0\rangle|P_1\rangle+\sqrt{p''_1}|\tilde{\Phi}''\rangle,
\label{eqn-49}
\end{eqnarray}
where the vector representative $|\varphi_1\rangle$ is defined by $|\varphi_1\rangle\propto\sqrt{r''}(\alpha|h\rangle+\beta|0\rangle)$ and $r''$ is a normalization constant. $|\tilde{\Phi}''\rangle$ is the vector representative of the failure state $\rho_{\tilde{\Phi}''}$ that should be annihilated by the projector $|P_1\rangle\langle P_1|$, i.e., ${\rm tr}(\rho_{\tilde{\Phi}''}|P_1\rangle\langle P_1|)=0$, and $p''_0$ is the success probability of recovering pure state $\rho_{\varphi_1}$.

Moreover, using the linearity of quantum operations, it follows from Eqs.(\ref{eqq-4}) and (\ref{eqq-5}) that
\begin{eqnarray}
&&U|h\rangle|\Sigma\rangle|P_0\rangle
\nonumber
\\
&\propto &
|\varphi_2\rangle|0\rangle
|P_1\rangle+a\sqrt{p_1}|\tilde{\Phi}_0\rangle
+b\sqrt{p'_1}|\tilde{\Phi}_1\rangle,
\label{eqn-50}
\end{eqnarray}
where the vector $|\varphi_2\rangle$ is given by $|\varphi_2\rangle=a\sqrt{p_0}|0\rangle+b\sqrt{p'_0}|\chi'\rangle$. From Eqs.(\ref{eqn-49}) and (\ref{eqn-50}), we get $a(\sqrt{\frac{p_0}{p_0'}}e^{i\theta}-\alpha)+\beta be^{i\theta}=\alpha\beta$ with $\theta\in[0,2\pi]$. This leads to the fixed values of $a$ and $b$ from $|a|^2+|b|^2=1$, where $p_0, p_0'$, $|\alpha|$ and $|\beta|$ are some fixed constants. This is a contradiction to the assumption of $\rho_{h}$. $\hfill{} \Box$

\subsection{Unified no-go theorems}

Theorem 1 presents the nonexistence of linear superposing of an unknown state and a fixed state in Hilbert space of a finite dimension. Similar results hold for different input states and output states.

\begin{thm}
Let $\alpha, \beta$ be complex constants satisfying $|\alpha|^2+|\beta|^2=1$ and $|\beta|\not=1$, $k$ be an integer, and $\rho_{\Phi}$ be a fixed pure state (known or unknown) on Hilbert space $\mathbb{H}^{n}$ with ${\rm dim}(\mathbb{H})\geq 2$. Then, there does not exist CP map ${\cal F}\in {\cal CP}(\mathbb{H}^k, \mathbb{H}^n)$ for all the states $\rho_{\psi}^{\otimes k}$ on $\mathbb{H}^k$ such that
\begin{eqnarray}
{\cal F}(\rho_{\psi}^{\otimes k})=\rho_{\Psi}
\label{eqn-51}
\end{eqnarray}
if $n, k$, $\beta$ and ${\cal F}$ satisfy one of the following conditions:
\begin{itemize}
\item[(i)] $n\geq 1$ and $0<|\beta|<1$;
\item[(ii)] $n>k$ and $\beta=0$;
\item[(iii)] $n<k$, $\beta=0$, the output state is restricted to $\rho_{\Psi}\rho_{0}^{\otimes k+n-m}$, and ${\cal F}$ is restricted to the unitary map;
 \end{itemize}
where the vector representative of $\rho_{\Psi}$ is given by $|\Psi\rangle\propto\sqrt{r}(\alpha|\psi\rangle^{\otimes n}+\beta|\Phi\rangle)$ and $r$ is a normalization constant.

\label{thm6}
\end{thm}

\begin{thm}
Let $\alpha, \beta$ be nonzero complex constants satisfying $|\alpha|^2+|\beta|^2=1$ and $|\beta|\not=1$, and $\rho_{\phi}$ be a fixed state (known or unknown) on Hilbert space $\mathbb{H}$ with ${\rm dim}(\mathbb{H})\geq 2$. Then there dose not exist CP map ${\cal F}\in {\cal CP}(\mathbb{H}^k, \mathbb{H}^n)$ for all the states $\rho_{\psi}^{\otimes k}$ on $\mathbb{H}^{k}$ such that
\begin{eqnarray}
{\cal F}(\rho_{\psi}^{\otimes k})= \rho_{\varphi}^{\otimes n}
\label{eqn-52}
\end{eqnarray}
if $n, k$, $\beta$ and ${\cal F}$ satisfy one of the following conditions:
\begin{itemize}
\item[(i)] $n\geq 1$ and $0<|\beta|<1$;
\item[(ii)] $n>k$ and $\beta=0$;
\item[(iii)]$n<k$, $\beta=0$, the output state is restricted to $\rho_{\varphi}^{\otimes n}\rho_{0}^{\otimes k-n}$, and ${\cal F}$ is restricted to unitary map;
\end{itemize}
where the vector representative of $\rho_{\varphi}$ is given by $|\varphi\rangle\propto\sqrt{r}(\alpha|\psi\rangle +\beta|\phi\rangle)$ and $r$ is a normalization constant.

\label{thm7}
\end{thm}

Theorem \ref{thm6} may be viewed as no-superposing of the product state $\rho_{\psi}^{\otimes k}$ and a fixed state $\rho_{\Phi}$ while Theorem \ref{thm7} may be viewed as the combination of the no-cloning and the no-encoding of an unknown pure state and a fixed state. Moreover, Theorem \ref{thm6} may be viewed as the no-cloning of $\rho_{\psi}^{\otimes n}$ with a fixed error term $|\Phi\rangle$ while Theorem \ref{thm7} may be viewed as the no-cloning of $\rho_{\psi}$ with a fixed error term $|\phi\rangle$. Theorems \ref{thm6} and \ref{thm7} are different except the special case of $\beta=0$, which leads to the no-cloning theorem \cite{1} ($n>k$) or the no-deleting theorem \cite{3} ($n<k$). The relationships among Theorems 2 and 3 and previous no-go theorems \cite{1,3,10} are schematically shown in Fig.1. From this Figure, the set defined by respective Theorem \ref{thm6} and \ref{thm7} includes the no-cloning theorem \cite{1}, the no-deleting theorem \cite{3}, Theorem 1, or the no-superposition theorem \cite{10} as a special case. Formally, we obtain the implications as follows:
\begin{eqnarray*}
\mbox{Theorem 2}&\Rightarrow &\mbox{Theorem 1}
\\
 &\Rightarrow& \mbox{The no-superposing Theorem},
\\
\mbox{Theorem 2}&\Rightarrow& \mbox{The no-cloning Theorem},
 \\
\mbox{Theorem 2}&\Rightarrow& \mbox{The no-deleting Theorem},
\\
\mbox{Theorem 3}&\Rightarrow &\mbox{Theorem 1}
\\
 &\Rightarrow& \mbox{The no-superposing Theorem},
\\
\mbox{Theorem 3}&\Rightarrow& \mbox{The no-cloning Theorem},
\\
\mbox{Theorem 3}&\Rightarrow& \mbox{The no-deleting Theorem.}
\end{eqnarray*}

\begin{figure}
\begin{center}
\resizebox{215pt}{115pt}{\includegraphics{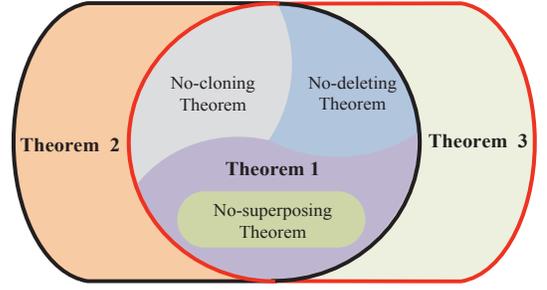}}
\end{center}
\caption{(Color online) Schematic relationships of different no-go theorems.  The set bounded by the black line denotes the results ensured by Theorem 2. The set bounded by the red line denotes the results ensured by Theorem 3.}
\end{figure}

{\it Proof of Theorem 2}. Since Theorem 1 is a special case of this Theorem, the proof of Theorem 1 for a general fixed state $\rho_{\phi}$ is included in the following proof.

When $\beta=0$ and ${\cal F}$ is restricted to the unitary map, Theorem 2 reduces to the no-deleting theorem. When $\beta=0$, Theorem 2 reduces to the no-cloning theorem for $n>k$. In the following, we only need to prove the result for $0<|\beta|<1$.

The proof of Theorem 1 suggests a method to prove this theorem. Now, we present another proof with simple evaluations. Assume that there exists a unitary map ${\cal U}$ with the matrix representative $U$ for all the states $\rho_{\psi}$ on $\mathbb{H}$ such that
\begin{eqnarray}
&&U|\varphi\rangle^{\otimes k}|\Sigma\rangle|P_0\rangle
\nonumber
\\
&\propto & \sqrt{p_0}|\Psi\rangle|0\rangle^{\otimes N+k-n}|P_1\rangle
+\sqrt{p_1}|\tilde{\Phi}\rangle,
\label{app5}
\end{eqnarray}
where the vector representative of $\rho_{\Psi}$ is given by $|\Psi\rangle\propto\sqrt{r}({\alpha}|\varphi\rangle^{\otimes n}+{\beta}|\Phi\rangle)$ and $r$ is a normalization constant. $|\Sigma\rangle$ is the vector representative of an ancillary state $\rho_{\Sigma}$ on Hilbert space $\mathbb{H}_A=\mathbb{H}^N$ with $N>n$, and $|P_0\rangle, |P_1\rangle$ are the respective vector representative of orthogonal states $\rho_{P_0}$ and $\rho_{P_1}$ on probe space $\mathbb{H}_P$ with ${\rm dim}(\mathbb{H}_P)\gg 2$. $|\tilde{\Phi}\rangle$ is the vector representative of the failure state $\rho_{\tilde{\Phi}}$ that should be annihilated by the projector $|P_1\rangle\langle P_1|$, i.e., ${\rm tr}(\rho_{\tilde{\Phi}}|P_1\rangle\langle P_1|)=0$.

For any state with the vector representative $e^{i\theta}|\varphi\rangle$, it follows from Eq.(\ref{app5}) that
\begin{eqnarray}
&&U(e^{i\theta}|\varphi\rangle^{\otimes k}|\Sigma\rangle|P_0\rangle)
\nonumber
\\
&\propto& \sqrt{p'_0} |\Psi'\rangle|0\rangle^{\otimes N+k-n}|P_1\rangle
+\sqrt{p'_1}|\tilde{\Phi}'\rangle,
\label{app6}
\end{eqnarray}
where the vector representative of $\rho_{\Psi'}$ is given by $|\Psi'\rangle\propto\sqrt{r'}(\alpha e^{in\theta}|\varphi\rangle^{\otimes n}+\beta|\Phi\rangle)$ and $r'$ is a normalization constant. $|\tilde{\Phi}'\rangle$ is the vector representative of a general failure state $\rho_{\tilde{\Phi}'}$ that should be annihilated by the projector $|P_1\rangle\langle P_1|$.

Note that $|\varphi\rangle\propto e^{i\theta}|\varphi\rangle$, i.e, the pure state $\rho_{e^{i\theta}\varphi}$ is physically undiscriminating from the pure state $\rho_{\varphi}$. From Eq.(\ref{app5}) we obtain
\begin{eqnarray}
U(e^{i\theta}|\varphi\rangle^{\otimes k}|\Sigma\rangle|P_0\rangle)
\propto  U|\varphi\rangle^{\otimes k}|\Sigma\rangle|P_0\rangle.
\label{app6'}
\end{eqnarray}
From Eqs.(\ref{app6}) and (\ref{app6'}) we obtain $\beta=0$ which is a contradiction to the assumption of $\beta$. $\hfill{} \Box$

The proof of Theorem \ref{thm7} is shown in Appendix B.

\section{Quantum transformation of pure states in a finite set}

Although the no-cloning theorem forbids perfectly cloning unknown states, Bu\v{z}ek and Hillery \cite{18} provided a cloning transformation for unknown qubit states such that the partial traces on the original qubit and on the cloned qubit have the same density matrix as $\rho =f\rho_{\psi}+ (1- f)\rho_{\psi^\bot}$ with the optimal fidelity $f=5/6$. Here, $\rho_{\psi^\bot}$ is the orthogonal state of the input state $\rho_{\psi}$. The cloning transformation of Bu\v{z}ek-Hillery triggered lots of investigations on the imperfect quantum cloning, see reviews \cite{19,Fan}.

Similar results hold for cloning pure states in a finite set. In fact, let $\mathbb{S}=\{\rho_{\psi_1}, \cdots, \rho_{\psi_m}\}$, where $\rho_{\psi_j}$ is on Hilbert space $\mathbb{H}$ with ${\rm dim}(\mathbb{H})\geq m$. There exists a CP map ${\cal F}\in {\cal CP}(\mathbb{S}\otimes \mathbb{H}_A, \mathbb{S}^2)$ \cite{13,14} for all the states in $\mathbb{S}$ such that ${\cal F}(\rho_{\psi_i}\rho_{\Sigma})=\rho_{\psi_i}\rho_{\psi_i}$ if the vector representatives of $\rho_{\psi_1}, \rho_{\psi_2}, \cdots, \rho_{\psi_n}$ are linearly independent, where $\rho_{\Sigma}$ is an ancillary state on Hilbert space $\mathbb{H}_A$ with ${\rm dim}(\mathbb{H}_A)\geq m$. Note that a CP map may be represented by a unitary evolution and post-selection of projective measurement. So, the cloning transformation ${\cal F}$ may be viewed as an imperfect cloning before performing the projective measurement. This scheme has been extended to probabilistically recover $\rho_{\psi_i}^{\otimes 2}$, $\rho_{\psi_i}^{\otimes 3}$, $\cdots$, $\rho_{\psi_i}^{\otimes M}$ for a large integer $M$ from the input states $\rho_{\psi_i}$, $i=1, 2, \cdots, m$ \cite{15}. Moreover, combined with the quantum deleting machine, i.e., $\rho_{\psi_i}\rho_{\psi_i}\mapsto \rho_{\psi_i}$, Feng {\it et al.} \cite{16} proved that there exists a CP map ${\cal F}^s$ for the same goal using the input states $\rho_{\psi_i}^{\otimes k}$, $i=1, 2, \cdots, m$. These optimal schemes are related to the quantum state discrimination \cite{CB,7b,CJW,MW}. In comparison to the imperfect cloning scheme \cite{18}, the error terms induced by these CP maps on a finite set are not restricted to the orthogonal states of the input states. The general transformation of the pure states in a finite set depends on the linear independence of the input states. The success probability is determined by a proper metric inequality \cite{14,15,16}.

In this section, our goal is to present the general transformations of a finite set of the pure states beyond previous probabilistic cloning or deleting \cite{13,14,15,16}.

\subsection{Hadamard product of matrices}

For completeness, some necessary definition and propositions of Hadamard product (Shur product) of matrices are firstly presented as follows \cite{17}.

\begin{df}
Let $A=[a_{ij}]_{m\times n}$ and $B=[b_{ij}]_{m\times n}$ be matrices in  matrix space $\mathbb{C}^{m\times n}$. The Hadamard product of $A$ and $B$ is defined as $A\circ B =[a_{ij}b_{ij}]_{m\times n}\in \mathbb{C}^{m\times n}$.

\label{df1}
\end{df}

\begin{pro}
The Hadamard product of matrices is commutative, associative and distributive.
\end{pro}

\begin{pro}
 The Hadamard product of two symmetric (Hermite) matrices is symmetric (Hermite).
\end{pro}

These Propositions may easily follow from their definitions. Moreover, the Hadamard product maintains the positive semidefinite property as the following theorem.

\begin{thm}
[{\bf Schur Product Theorem}] \cite{17}. The Hadamard product of two positive-semidefinite matrices (or positive-definite) is also positive-semidefinite (or positive-definite).
\label{sp}
\end{thm}

\begin{lm}
Let $A$ be a positive definite matrix in matrix space $\mathbb{C}^{n\times n}$ and $B$ be a Hermite matrix in the same space. Then $A-B$ is positive definite if
\begin{eqnarray}
\|B\|_2<\|A^{-1}\|^{-1}_{2}
\label{eqn-7}
\end{eqnarray}
where $\|\cdot\|_2$ is the induced norm from the Euclidean metric.

\label{lm-1}
\end{lm}

{\it Proof of Lemma 1}. For a positive definite matrix $A$ \cite{17}, all the eigenvalues $\lambda_1, \lambda_2, \cdots, \lambda_n$ are positive constants. From the definition of the induced norm, we have $\|A\|_2=\max\{\lambda_1, \lambda_2, \cdots, \lambda_n\}$ and $\|A^{-1}\|_2^{-1}=\min\{\lambda_1, \lambda_2, \cdots, \lambda_n\}$ where $A^{-1}$ denotes the inverse matrix of $A$. For a Hermite matrix $B$, all the eigenvalues $\lambda'_1, \lambda'_2, \cdots, \lambda'_n$ are nonnegative and satisfy $\|B\|_2=\max\{\lambda'_1, \lambda'_2, \cdots, \lambda'_n\}$. If $A-B$ is negative semi-definite, there exists a nonzero vector $v\in\mathbb{C}^n$ such that $v(A-B)v^\dag\leq 0$. It means that
\begin{eqnarray*}
\|A^{-1}\|_2^{-1}\leq vAv^\dag\leq vBv^\dag\leq \|B\|_2,
\end{eqnarray*}
which contradicts the inequality of (\ref{eqn-7}). $\hfill{} \Box$

\subsection{General quantum transformation of finite set}

In this section, the imperfect and perfect transformation of the pure states in a finite set will be unified. In detail, let $S$ be a finite set defined by $\mathbb{S}=\{\rho_{\psi_1}^{\otimes k}, \rho_{\psi_2}^{\otimes k}, \cdots, \rho_{\psi_m}^{\otimes k}\}\subset\mathbb{H}^k$, where ${\rm dim}(\mathbb{H})\geq 2$ and $k$ is a positive integer.

\begin{thm}
Let $\alpha_{ij}$ and $\beta_{ij}$ be complex constants satisfying $|\alpha_{ij}|^2+|\beta_{ij}|^2=1$ and $\sum_{j=1}^{M+k}|\alpha_{ij}|^2\not=0$, and $\rho_{{\Phi}_{ij}}$ be a known pure state on Hilbert space $\mathbb{H}^{j}$, then there exists a CP map ${\cal F}_j\in {\cal CP}(\mathbb{H}^{k}, \mathbb{H}^{j})$ for all the states on $\mathbb{S}$ such that
\begin{eqnarray}
{\cal F}_j(\rho_{\psi_i}^{\otimes k})=\rho_{\varphi_{ij}}
\label{eqn-9'}
\end{eqnarray}
iff the vector representatives of $\rho_{\psi_1}, \rho_{\psi_2}, \cdots, \rho_{\psi_m}$ are linearly independent, where the vector representative of $\rho_{\varphi_{ij}}$ is given by $|\varphi_{ij}\rangle\propto\sqrt{r_{ij}}(\alpha_{ij}
|\psi_i\rangle^{\otimes j}+\beta_{ij}|{\Phi}_{ij}\rangle)$ and $r_{ij}$ is a normalization constant, $i=1, 2, \cdots, m$; $j=1, 2, \cdots, M+k$, and $M$ is a positive integer.

 \label{thm1}
\end{thm}

In Theorem 5, the CP maps exist for $\mathbb{S}$ when the vector representatives of $\mathbb{S}$ are linearly independent. If $\mathbb{S}$ is known, a CP map \cite{7} exists for recovering the states in $\mathbb{S}$ from a new set $\{\rho_{\psi_1}, \rho_{\psi_2}, \cdots, \rho_{\psi_m}\}$. Hence, our condition will be replaced by the linear independence of the vectors $|\psi_1\rangle, |\psi_2\rangle, \cdots, |\psi_m\rangle$ for a large $k$. This is because that the vector representatives of $\mathbb{S}$ should be linearly independent when $k$ is large using the Gershgorin circle theorem \cite{Gers} for its Gram matrix $A^{\circ k}$.

Before we present the proof, some useful results may be firstly followed from Theorem 5. When $\beta_{ij}=0$ for all $i=1, 2, \cdots, m$ and $j=1, 2, \cdots, M+k$, the general map in Eq.(\ref{eqn-9'}) reduces to the unified cloning \cite{16}. If $\beta_{ij}=0$ holds for all $i=1, 2, \cdots, m$ and $j=1, 2, \cdots, k_1$ with $1\leq k_1<M+k$, Theorem \ref{thm1} reduces to a simultaneous perfect and imperfect cloning as follows.

\begin{cor}
Let $\alpha_{ij},\beta_{ij}$ be complex constants satisfying $|\alpha_{ij}|^2+|\beta_{ij}|^2=1$ and $\sum_{j=1}^{M+k}|\alpha_{ij}|^2\not=0$, $k_1$ be an integer satisfying $1\leq k_1< M+k$, and $\rho_{{\Phi}_{ij}}$ be a known pure state on Hilbert space $\mathbb{H}^{j}$, then there exists a CP map ${\cal F}_j\in {\cal CP}(\mathbb{H}^{k}, \mathbb{H}^{k_1+j})$ for all the states in $\mathbb{S}$ such that
\begin{eqnarray}
{\cal F}_j(\rho_{\psi_i}^{\otimes k})=\rho_{\psi_i}^{\otimes k_1}\rho_{\varphi_{ij}}
\label{eqn-25}
\end{eqnarray}
iff the vector representatives of $\rho_{\psi_1}, \rho_{\psi_2}, \cdots, \rho_{\psi_m}$ are linearly independent, where the vector representative of $\rho_{\varphi_{ij}}$ is given by $|\varphi_{ij}\rangle\propto\sqrt{r_{ij}}(\alpha_{ij}
|\psi_i\rangle^{\otimes j}+\beta_{ij}|{\Phi}_{ij}\rangle)$ and $r_{ij}$ is a normalization constant, $i=1, 2, \cdots, m$; $j=1, 2, \cdots, L$,  $L=M+k-k_1$ and $M$ is a positive integer.
\label{cor3}
\end{cor}

From Eq.(\ref{eqn-25}), one may simultaneously produce $k_1$ perfect copies and $j$ imperfect copies of $\rho_{\psi_i}$ when $k_1>k$. Moreover, one may delete $k-k_1$ copies and clone $j$ imperfect copies of $\rho_{\psi_i}$ when $k_1<k$. The imperfect terms $|{\Phi}_{ij}\rangle$ ($i=1, 2, \cdots, m; j=1, 2, \cdots, L$) are independent of the input states. When $k_1<k$, one may keep $k_1$ copies of $\rho_{\psi_i}$ being unchanged and perform a CP map ${\cal F}$ on the remained $k-k_1$ copies from Theorem \ref{thm1}.

Note that for arbitrary complex constants $\hat{\alpha}_{ij}, \hat{\beta}_{ij}$ satisfying $|\hat{\alpha}_{ij}|^2+|\hat{\beta}_{ij}|^2=1$, and normalized vectors $|\psi_i\rangle, |{\phi}_{ij}\rangle \in \mathbb{C}^K$ with $K\geq 2$, there exist complex constants $\alpha_{ij}, \beta_{ij}$ and a normalized vector $|{\Phi}_{ij}\rangle\in \mathbb{C}^{nK}$ such that $\alpha_{ij}|\psi_i\rangle^{\otimes n}+\beta_{ij}|{\Phi}_{ij}\rangle \propto (\hat{\alpha}_{ij}|\psi_i\rangle +\hat{\beta}_{ij}|{\phi}_{ij}\rangle)^{\otimes n}$ with $|{\alpha}_{ij}|^2+|{\beta}_{ij}|^2=1$. From similar proof, this fact may be generally reformed as follows.

\begin{cor}
Let $\alpha_{ij},\beta_{ij}$ be complex constants satisfying $|\alpha_{ij}|^2+|\beta_{ij}|^2=1$ and $\sum_{j=1}^{M+k}|\alpha_{ij}|^2\not=0$, $k_1$ be an integer satisfying $1\leq k_1< M+k$, and $\rho_{\phi_{ij}}$ be a known pure state on $\mathbb{H}$, then there exists a CP map ${\cal F}_j\in {\cal CP}(\mathbb{H}^{k}, \mathbb{H}^{k_1+j})$ for all the states in $\mathbb{S}$ such that
\begin{eqnarray}
{\cal F}_j(\rho_{\psi_i}^{\otimes k})=\rho_{\psi_i}^{\otimes k_1}\rho_{\varphi_{ij}}^{\otimes j}
\label{eqn-26}
\end{eqnarray}
iff the vector representatives of $\rho_{\psi_1}$, $\rho_{\psi_2}$, $ \cdots, \rho_{\psi_m}$ are linearly independent, where the vector representative of $\rho_{\varphi_{ij}}$ is defined by $|\varphi_{ij}\rangle\propto\sqrt{r_{ij}}(\alpha_{ij}|\psi_i\rangle
+\beta_{ij}|{\phi}_{ij}\rangle)$ and $r_{ij}$ is a normalization constant, $i=1, 2, \cdots, m$, $j=1, 2, \cdots, L$ and $L=M+k-k_1$.

\label{cor-4}
\end{cor}

The no-superposing theorem \cite{12} forbids superposing an unknown state and its orthogonal state in Hilbert space of a finite dimension. However, Corollary 2 suggests a perfect superposing for unknown states in a finite set. If we choose complex constants $\alpha_{ij}, \beta_{ij}$ and a normalized vector $|\phi_{ij}\rangle\in \mathbb{C}^{K}$ with $K\geq 2$ such that $\alpha_{ij}|\psi_i\rangle+\beta_{ij}|{\phi}_{ij}\rangle\propto |\psi^\bot_i\rangle$ which is the orthogonal to $|\psi_i\rangle$, the Corollary \ref{cor-4} leads to the following result with similar proof.

\begin{cor}
Let $\rho_{\phi_{ij}}$ be a known pure state on $\mathbb{H}$, then there exists a CP map ${\cal F}_j\in {\cal CP}(\mathbb{H}^{k},\mathbb{H}^{k_1+j})$ for all the states in $\mathbb{S}$ such that
\begin{eqnarray}
{\cal F}_j(\rho_{\psi_i}^{\otimes k})=\rho_{\psi_i}^{\otimes k_1}\rho_{\psi^\bot_i}^{\otimes j}
\label{eqn-27}
\end{eqnarray}
iff the vector representatives of $\rho_{\psi_1}$, $\rho_{\psi_2}$, $\cdots, \rho_{\psi_m}$ are linearly independent, where $i=1, 2, \cdots, m$; $j=1, 2, \cdots, L$, $L=M+k-k_1$ and $M$ is a positive integer.

\label{cor5}
\end{cor}

{\it Proof of Theorem \ref{thm1}}. The proof consists of two parts. First, we prove that if there exists a CP map ${\cal F}_j$ satisfying Eq.(\ref{eqn-9'}), then the vector representatives of $\rho_{\psi_1}, \rho_{\psi_2}, \cdots, \rho_{\psi_m}$ are linearly independent. These CP maps will be presented in a unified form with a nontrivial probability in the following. In fact, from Stinespring's Theorem \cite{Stin}, assume that there exists a unitary map ${\cal U}$ with matrix representative $U$, an ancillary state with the vector representative $|\Sigma\rangle$ on $\mathbb{H}_A$, $N+1$ orthogonal states with the vector representatives $|P_0\rangle, |P_1\rangle, \cdots, |P_N\rangle$ on $\mathbb{H}_P$ such that
 \begin{eqnarray}
U|\psi_i\rangle^{\otimes k}|\Sigma\rangle|P_0\rangle
&\propto&\sum_{j=1}^{M+k} \sqrt{p_{ij}}|\varphi_j\rangle|0\rangle^{\otimes M+k-j}|P_j\rangle
\nonumber
\\
&&+\sum_{s=M+k+1}^N \sqrt{q_{is}}|\tilde{\Phi}_{is}\rangle|P_s\rangle
\label{eqn-9}
\end{eqnarray}
where $\mathbb{H}_A=\mathbb{H}^M$ be an ancillary Hilbert space with $M>k$, and $\mathbb{H}_P$ be a probe space with ${\rm dim}(\mathbb{H}_P)\geq N+1>M+k$. $|{\Phi}_{ij}\rangle$ is the vector representative of normalized state $\rho_{\Phi_{ij}}$ that is independent of the input states, $i=1, 2, \cdots, m; j=1, 2, \cdots, M+k$. $p_{ij}$ is the success probability of producing $n$ imperfect copies of $\rho_{{\psi_i}}$. $|\tilde{\Phi}_{is}\rangle$ is the vector representative of a general failure state $\rho_{\tilde{\Phi}_{is}}$ on $\mathbb{H}^k\otimes \mathbb{H}_A$  with the probability $q_{is}$ that satisfies $\sum_{j=1}^{M+k}p_{ij}+\sum_{s=M+k+1}^Nq_{is}=1$, $i=1, 2, \cdots, M+k$, and $s=M+k+1, M+k+2, \cdots, N$. Each CP map ${\cal F}_j$ is defined by ${\cal F}_j=|0\rangle^{M+k-j}\langle0|{}^{M+k-j}\circ |P_j\rangle\langle P_j|\circ {\cal U}$, $j=1, 2, \cdots, M+k$.

Assume that there exist a vector $|\psi_j\rangle^{\otimes k}$ and constants $c_{1j}, c_{2j}, \cdots, c_{mj}$ such that $|\psi_j\rangle^{\otimes k}=\sum_{i=1}^mc_{ij}|\psi_i\rangle^{\otimes k}$. From Eq.(\ref{eqn-9}) we have
\begin{eqnarray}
 U|\Psi_j\rangle
&\propto&\sum_{s=1}^{M+k} \sqrt{p_{js}}|\varphi_s\rangle|0\rangle^{\otimes M+k-s}|P_s\rangle
\nonumber
\\
&&+\sum_{t=M+k+1}^N \sqrt{q_{jt}}|\tilde{\Phi}_{jt}\rangle|P_t\rangle
\label{eqn-10}
\end{eqnarray}
where the vector representative of $\rho_{\varphi_s}$ is given by $|\varphi_s\rangle\propto\sqrt{r_{is}}(\alpha_{is}
|\psi_i\rangle^{\otimes s}+\beta_{is}|{\Phi}_{is}\rangle)$ and the vector $|\Psi_j\rangle=|\psi_j\rangle^{\otimes k}|\Sigma\rangle|P_0\rangle$. Moreover, from the linearity of quantum operations, we obtain that
\begin{eqnarray}
U|\Psi_j\rangle
&\propto&
\sum_{s=1}^{M+k}\sum_{i=1}^m\sqrt{c_{ij}} \sqrt{p_{is}}|\varphi_s\rangle|0\rangle^{\otimes M+k-s}|P_s\rangle
\nonumber
\\
&&+\sum_{t=M+k+1}^N\sum_{i=1}^m\sqrt{c_{ij}} \sqrt{q_{it}}|\tilde{\Phi}_{it}\rangle|P_t\rangle
\label{eqn-11}
\end{eqnarray}
Note that Eqs.(\ref{eqn-10}) and (\ref{eqn-11}) must be equal to each other for all the states $\rho_{\psi_i}$ (up to a phase factor), $i=1, 2,\cdots, m$. It follows that $\sum_{i=1}^m\sqrt{c_{ij} p_{is}}|\varphi_s\rangle\propto\sqrt{p_{js}}|\varphi_s\rangle$ for all the integers $s$ satisfying $1\leq s\leq M+k$. These equalities imply that $c_{jj}=1$ and $c_{ij}=0$ for $i\not=j$ because the vector $|{\Phi}_{ij}\rangle$ is independent of input states. So, the vector representatives of $\rho_{\psi_1}, \rho_{\psi_2}, \cdots, \rho_{\psi_m}$ are linearly independent.

Now, we need to prove the existence of the unitary map satisfying  Eq.(\ref{eqn-9}). For the input states $\rho_{\psi_i}^{\otimes k}$ and $\rho_{\psi_j}^{\otimes k}$, taking the inner product of the representative vectors of the output states \cite{56}, from Eq.(\ref{eqn-9}) we have
\begin{eqnarray}
a_{ij}^k
&=&\sum_{s=1}^{M+k}\sqrt{p_{is}}(
\sqrt{r_{is}}\alpha_{is}a^s_{ij}\alpha^*_{js}\sqrt{r_{js}}
\nonumber
\\
&&+\alpha_{is}\lambda_{ij,s}\beta^*_{js}
+\beta_{is}\lambda^*_{ji,s}\alpha^*_{js}
+\beta_{is}\beta^*_{js})\sqrt{p_{js}}
\nonumber
\\
&&
+\sum_{\ell=M+k+1}^N\sqrt{q_{i\ell}q_{j\ell}},
\label{eqn-12}
\end{eqnarray}
where $a_{ij}:=\langle \psi_i|\psi_j\rangle$ and $\lambda_{ij,s}:=\langle \psi_i|{}^{\otimes s}|{\Phi}_{js}\rangle$. Briefly, we obtain a matrix equation using Definition 1 as follows:
\begin{eqnarray}
A^{\circ k}=\sum_{s=1}^{M+k}G_s(\Lambda_sA^{\circ s}\Lambda_s^\dag+C_s ) G_s^\dag
+\sum_{\ell=M+k+1}^NQ_\ell,
\label{eqn-13}
\end{eqnarray}
where $A^{\circ t}$ denotes the $t$-fold Hadamard power of matrix $A=[a_{ij}]_{m\times m}$, $C_s=[\alpha_{is}\lambda_{ij,s}\beta^*_{js}+
\beta_{is}\lambda^*_{ji,s}\alpha^*_{js}+\beta_{is}\beta^*_{js}]_{m\times m}$, $Q_\ell=[\sqrt{q_{i\ell} q_{j\ell}}]_{m\times m}$, and diagonal matrices $G_s=G_s^\dag={\rm diag}(\sqrt{p_{1s}}, \sqrt{p_{2s}}, \cdots, \sqrt{p_{ms}})$ and $\Lambda_s={\rm diag}(\alpha_{1s}\sqrt{r_{1s}}, \alpha_{2s}\sqrt{r_{2s}}, \cdots, \alpha_{ms}\sqrt{r_{ms}})$.

From Lemma 1 \cite{14}, it is sufficient to prove Eq.(\ref{eqn-13}) with physically available matrices $Q_{M+k+1}$, $Q_{M+k+2}, \cdots, Q_{N}$ for the existence of the unitary map in Eq.(\ref{eqn-9}). The matrix $A^{\circ k}$ is positive definite from Theorem 4 because the vectors $|\psi_1\rangle$, $|\psi_2\rangle$, $\cdots, |\psi_m\rangle$ are linearly independent. The matrices $\sum_{s=1}^{M+k}\Lambda_sA^{\circ s}\Lambda_s^\dag$ and $C_s$ are Hermite. From Lemma 1, when the efficiency matrices $G_s (s=1,2, \cdots, M+k)$ satisfy the following inequality
\begin{eqnarray}
\|\sum_{s=1}^{M+k}G_sB_s G_s^\dag\|_2<\|(A^{\circ k})^{-1}\|_2^{-1},
\label{eqn-14}
\end{eqnarray}
the matrix $A^{\circ k}-\sum_{s=1}^{M+k}G_sB_s G_s^\dag$
is positive definite, where $B_s=\Lambda_sA^{\circ s}\Lambda_s^\dag +C_s$. So, the Hermite matrix $A^{\circ k}-\sum_{s=1}^{M+k}G_sB_s G_s^\dag$ can be diagonalized by a unitary matrix $V$ as follows:
\begin{eqnarray}
V(A^{\circ k}-\sum_{s=1}^{M+k}G_sB_s G_s^\dag)V^\dag
={\rm diag}(\gamma_{1}, \gamma_{2}, \cdots, \gamma_{m})
\label{eqn-15}
\end{eqnarray}
where $\gamma_{1}, \gamma_{2}, \cdots, \gamma_{m}$ denote all the eigenvalues of $A^{\circ k}-\sum_{s=1}^{M+k}G_sB_s G_s^\dag$ and $\gamma_{j}>0$, $j=1, 2, \cdots, m$. Define $Q_\ell=V^\dag{\rm diag}(x_{1\ell}, x_{2\ell}, \cdots, x_{m\ell})V$ where $x_{j\ell}$ is nonnegative and satisfies $\sum_{\ell=M+k+1}^Nx_{j\ell}=\gamma_{j}$, $j=1, 2, \cdots, m$. Hence, the matrices $Q_{M+k+1}$, $Q_{M+k+2}$, $\cdots, Q_N$ are positive definite from Theorem \ref{sp} and then physically realizable.

Before completing the proof, we need to prove the inequality (\ref{eqn-14}). Using the triangle inequality, it easily follows that $\|\sum_{s=1}^{M+k}G_sB_s G_s^\dag\|_2\leq \sum_{s=1}^{M+k}\|G_s\|^2_2\times\|B_s\|_2$. So, there exist nonzero matrices $G_s$ ($s=1, 2, \cdots, M+k$) satisfying Eq.(\ref{eqn-14}) when they satisfy the following inequality
\begin{eqnarray}
\sum_{s=1}^{M+k}\|G_s\|^2_2\times\|B_s\|_2
<\|(A^{\circ k})^{-1}\|_2^{-1}
\label{eqn-16}
\end{eqnarray}
which is easily guaranteed. This completes the proof.  $\hfill{} \Box$

The inequality (\ref{eqn-14}) or its weak form (\ref{eqn-16}) has presented an implicit bound of the efficiency in terms of the matrix norm. An entry-pair inequality is obtained from Eq.(\ref{eqn-13}) using the state metric \cite{12} as follows
\begin{eqnarray}
D_{ij,k}&\geq&
\sum_{s=1}^{M+k}p_{ij,s}[\alpha_{ij,s}(r_{ij,s}D^2_{ij,s}-2r_{ij,s}+2)
\nonumber
\\
&&-2\max\{|\lambda_{ij,s}|,|\lambda_{ji,s}|\})
\label{eqn-21}
\end{eqnarray}
where ${p}_{ij,s}=(p_{is}+p_{js})/2$, $\alpha_{ij,s}=|\alpha_{is}\alpha_{js}|$, $\beta_{ij,s}=|\beta_{is}\beta_{js}|$, ${r}_{ij,s}=\sqrt{r_{is}r_{js}}$, and  $D_{ij,t}=2(1-|a_{ij}|^t)$ with $t=k$ or $s$. The proof is shown in Appendix C. This inequality has generalized the bound of the probabilistic cloning \cite{7} or the state discrimination \cite{CB}. It also reduces to the bound of the probabilistic cloning in a unified form \cite{12,14,15,16} for special constants of $\lambda_{ij,s}$, $\alpha_{ij}$ and $\beta_{ij}$.

{\it Remark}. From the cloning scheme \cite{14}, the final states belong to a finite set of $\cup_{j=1}^{M+k}\mathbb{S}_j$, where $\mathbb{S}_j=\{\rho_{\psi_1}^{\otimes j}, \rho_{\psi_2}^{\otimes j}, \cdots, \rho_{\psi_m}^{\otimes j}\}$. Similar results hold for $\mathbb{S}_k$  \cite{16}. Moreover, from Theorem \ref{thm1}, the final states belong to a new set $\cup_{j=1}^{M+k}\tilde{\mathbb{S}}_j$, where $\tilde{\mathbb{S}}_j$ denotes an infinite vector set $\{\alpha_{ij}|\psi_i\rangle^{\otimes j}+\beta_{ij}|{\Phi}_{ij}\rangle, |\alpha_{ij}|^2+|\beta_{ij}|^2=1, i=1, 2, \cdots, m\}$ with fixed vectors $|{\Phi}_{ij}\rangle\in \mathbb{H}^j$. In geometry, each subset $\tilde{\mathbb{S}}_j$ may be viewed as a ``compressed unit sphere" with two axes of $|\psi_i\rangle^{\otimes j}$ and $|{\Phi}_{ij}\rangle$, see a schematic example shown in Fig.2.

\begin{figure}
\begin{center}
\resizebox{240pt}{165pt}{\includegraphics{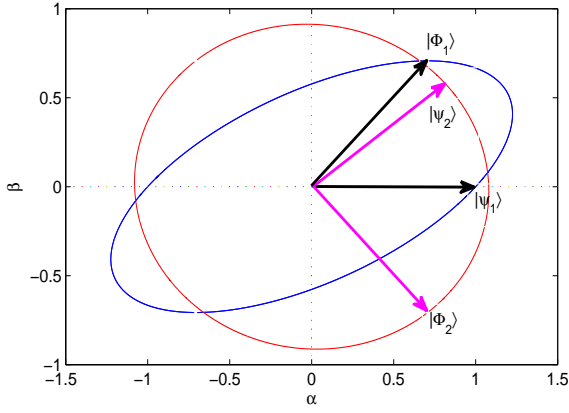}}
\end{center}
\caption{(Color online) Schematic representatives of $\tilde{\mathbb{S}}_1$ and $\mathbb{S}_1$ in $\mathbb{C}^2$. Here, the vectors $|\psi_1\rangle=|0\rangle$ and $|\Phi_1\rangle=\frac{1}{\sqrt{2}}(|0\rangle+|1\rangle)$, $|\psi_2\rangle=\sqrt{\frac{2}{3}}|0\rangle+\sqrt{\frac{1}{3}}|1\rangle$ and $|\Phi_2\rangle=\frac{1}{\sqrt{2}}(|0\rangle-|1\rangle)$. $\mathbb{S}_1=\{\rho_{\psi_1},\rho_{\psi_2}\}$ is a set of two points. $\tilde{\mathbb{S}}_1=\{\alpha_i|\psi_i\rangle+\beta_i|\Phi_i\rangle, |\alpha_i|^2+|\beta_i|^2=1, i=1, 2\}$ is a set consisting of all the points on two circles.}
\end{figure}

\section{Superposition of pure states in restricted sets}

Recent schemes show that one may superpose unknown states with a fixed projector onto a known state \cite{10}. Our goal in this section is to propose new superposing schemes for the input states in different restricted sets.

\subsection{Pure states in a finite set}

In this subsection, we investigate the superposed states belonging to a finite set $\mathbb{S}=\{\rho_{\psi_1}, \rho_{\psi_2}, \cdots, \rho_{\psi_m}\}\subset\mathbb{H}$ with ${\rm dim}(\mathbb{H})\geq m$. Note that the linear independence of the input states is a special case of the fixed overlap \cite{10,Form}. Unfortunately, the scheme proposed by Oszmaniec et al \cite{10} is only sufficient. New schemes present the necessity of the superposition scheme when the input states belong to a finite set.

\begin{thm}
Let $\alpha_{ij}, \beta_{ij}$ be nonzero complex constants satisfying $|\alpha_{ij}|^2+|\beta_{ij}|^2=1$, then there exists a CP map ${\cal F}\in {\cal CP}(\mathbb{H}^2, \mathbb{H})$ for all the states in $\mathbb{S}$ such that
\begin{eqnarray}
{\cal F}(\rho_{\psi_i}\rho_{\psi_j})=\rho_{\varphi_{ij}}
\label{eqn-54'}
\end{eqnarray}
iff the vector representatives of $\rho_{\psi_1}, \rho_{\psi_2}, \cdots, \rho_{\psi_m}$ are linearly independent, where the vector representative of $\rho_{\varphi_{ij}}$ is defined by $|\varphi_{ij}\rangle\propto\sqrt{r_{ij}}
(\alpha_{ij}|\psi_i\rangle+\beta_{ij}|\psi_j\rangle)$ and $r_{ij}$ is a normalization constant, $i,j=1,2, \cdots, m$.

\label{thm8}
\end{thm}

{\it Proof of Theorem \ref{thm8}}. Note that the vector representatives $|\psi_1\rangle|\psi_1\rangle$, $|\psi_1\rangle|\psi_2\rangle$, $\cdots, |\psi_m\rangle|\psi_m\rangle$ are linearly dependent iff the vectors $|\psi_1\rangle, |\psi_2\rangle, \cdots, |\psi_m\rangle$ are linearly dependent. Similar to the proof of Theorem 5, the necessity is easily followed from the linearity of quantum operations. Now we need to prove the existence of the CP map. Assume there exists a CP map ${\cal F}$ satisfying Eq.(\ref{eqn-54'}). From Stinespring's Theorem \cite{Stin}, there exists a unitary map with the matrix representative $U$, an ancillary state with vector representative $|\Sigma\rangle$ on Hilbert space $\mathbb{H}_A=\mathbb{H}^M$ with $M>m$, $3$ orthogonal states with the vector representatives $|P_0\rangle, |P_1\rangle, |P_2\rangle$ on $\mathbb{H}_P$ such that
\begin{eqnarray}
&& U|\psi_i\rangle|\psi_j\rangle|\Sigma\rangle|P_0\rangle
\nonumber
\\
&\propto &\sqrt{p_{ij}}|\varphi_{ij}\rangle|0\rangle|P_1\rangle
+\sqrt{q_{ij}}|\Phi_{ij}\rangle|P_2\rangle
\label{eqn-54}
\end{eqnarray}
Here, $p_{ij}$ is the success probability of recovering a superposed state $\rho_{\varphi_{ij}}$ with the vector representative $|\varphi_{ij}\rangle\propto\sqrt{r_{ij}}(\alpha_{ij}|\psi_i\rangle+\beta_{ij}|\psi_j\rangle)$, and satisfies $p_{ij}=p_{ij}$ which corresponds to the swapping of the input states. $|\Phi_{ij}\rangle$ is the vector representative of a normalized failure state $\rho_{\Phi_{ij}}$ on $\mathbb{H}\otimes \mathbb{H}_A$ with the probability $q_{ij}$ that satisfies $q_{ij}+p_{ij}=1$, $i,j=1, 2, \cdots, m$.

For the input states $\rho_{\psi_{i_1}}\rho_{\psi_{i_2}}$ and $\rho_{\psi_{j_1}}\rho_{\psi_{j_2}}$, taking the inner product of the vector representatives of the output states from Eq.(\ref{eqn-54}), we can obtain
\begin{eqnarray}
a_{i_1j_1}a_{i_2j_2}
&=&\sqrt{p_{i_1i_2}r_{i_1i_2}}h_{i_1i_2,j_1j_2}
\sqrt{p_{j_1j_2}r_{j_1j_2}}
\nonumber
\\
&&
+\sqrt{q_{i_1i_2}q_{j_1j_2}}b_{i_1i_2,j_1j_2}
\label{eqn-55}
\end{eqnarray}
where $a_{ks}=\langle\psi_k|\psi_s\rangle$, $h_{i_1i_2,j_1j_2}=\langle \Phi_{ij}|\Phi_{j_1j_2}\rangle$, and $h_{i_1i_2,j_1j_2}=\alpha^*_{i_1}\alpha_{j_1}a_{i_1j_1}
+\beta^*_{i_2}\beta_{j_2}a_{i_2j_2}+\alpha^*_{i_1}\beta_{j_2}a_{i_1j_2}
+\beta^*_{i_2}\alpha_{j_1}a_{i_2j_1}$. It may be briefly described as a matrix equation as follows
\begin{eqnarray}
A=\Lambda H \Lambda^\dag+ Q,
\label{eqn-56}
\end{eqnarray}
where the matrices $A, H, Q$ are respectively defined by
\begin{eqnarray*}
&&A=\left[a_{i_1j_1}a_{i_2j_2}\right]_{m^2\times m^2},
\\
&&\Lambda={\rm diag}(\sqrt{p_{11}}, \sqrt{p_{12}}, \cdots, \sqrt{p_{mm}})_{m^2\times m^2},
\\
&&H=\left[
\begin{array}{ll}
\sqrt{r_{i_1i_2}r_{j_1j_2}}h_{i_1i_2,j_1j_2}
\end{array}
\right]_{m^2\times m^2},
 \\
&&Q=[\sqrt{q_{i_1i_2}q_{j_1j_2}}b_{i_1i_2,j_1j_2}]_{m^2\times m^2},
\end{eqnarray*}
and the rows or columns of the matrices $A, H, Q$ are represented by two-bit series $i_1i_2$ and $j_1j_2$, respectively.

It is sufficient to prove Eq.(\ref{eqn-56}) with physically realizable matrix $Q$ for the existence of the unitary map in Eq.(\ref{eqn-54}). $A$ is positive definite and $\Lambda H \Lambda^\dag$ is Hermite. From Lemma \ref{lm-1}, $A-\Lambda H \Lambda^\dag$ is positive definite when $\Lambda$ satisfies
\begin{eqnarray}
\|\Lambda H \Lambda^\dag\|_2<\|A^{-1}\|_2^{-1}.
\label{eqn-57}
\end{eqnarray}
Therefore, there exists a unitary matrix $V$ such that
\begin{eqnarray}
V(A-\Lambda H \Lambda^\dag)V^\dag={\rm diag}(\lambda_{1}, \lambda_{2}, \cdots, \lambda_{m^2}),
\label{eqn-58}
\end{eqnarray}
where $\lambda_{1}, \lambda_{2}, \cdots, \lambda_{m^2}$ denote all the eigenvalues of $A-\Lambda H \Lambda^\dag$ and are positive. Denote $Q=V^\dag {\rm diag}(\lambda_{1}$, $\lambda_{2}, \cdots, \lambda_{m^2})V$. So, $Q$ is positive from Theorem \ref{sp} and then physically realizable. This completes the proof. $\hfill{} \Box$

The bound of the efficiency may be proved from Eq.(\ref{eqn-55}) as follows \begin{eqnarray}
D_{i_1j_1}D_{i_2j_2}
&\leq&\frac{2}{3} p_{i_1i_2,j_1j_2}[r_{i_1i_2,j_1j_2}(16-D_{i_1j_1}
\nonumber
\\
&&-D_{i_2j_1}-D_{i_1j_2}-D_{i_2j_2})-6]
\nonumber
\\
&&+4D_{i_1j_1}+4D_{i_2j_2},
\label{eqn-37'}
\end{eqnarray}
where $r_{i_1i_2,j_1j_2}=\sqrt{r_{i_1i_2}r_{j_1j_2}}$, $p_{i_1i_2,j_1j_2}=(p_{i_1i_2}+p_{j_1j_2})/2$ and $D_{ij}=2(1-|a_{ij}|)$. This is a new bound of the average efficiency $p_{i_1i_2,j_1j_2}$ in comparison to the bound in Eq.(\ref{eqn-21}) because Theorem 6 cannot be proved from Theorem 5.

Combined with the probabilistic cloning or deleting of Theorem \ref{thm1}, we obtain a general theorem as follows:

\begin{thm}
Let $\alpha_{ij},\beta_{ij}$ be nonzero complex constants satisfying $|\alpha_{ij}|^2+|\beta_{ij}|^2=1$, then there exists a CP map ${\cal F}\in {\cal CP}(\mathbb{H}^k\otimes \mathbb{H}^k, \mathbb{H}^s)$ for all the states in $\mathbb{S}=\{\rho_{\psi_1}^{\otimes k}, \rho_{\psi_2}^{\otimes k},\cdots, \rho_{{\psi_m}}^{\otimes k}\}\subset \mathbb{H}^k$ such that
\begin{eqnarray}
{\cal F}(\rho_{\psi_i}^{\otimes k}\rho_{{\psi_j}}^{\otimes k})=\rho_{\varphi_{ij,s}}
\label{new}
\end{eqnarray}
iff the vector representatives of $\rho_{\psi_1}^{\otimes k}, \rho_{\psi_2}^{\otimes k}, \cdots, \rho_{\psi_m}^{\otimes k}$ are linearly independent, where the vector representative of $\rho_{\varphi_{ij,s}}$ is defined by $|\varphi_{ij,s}\rangle\propto\sqrt{r_{ij,s}}(\alpha_{ij}|\psi_i\rangle^{\otimes s}+\beta_{ij}|\psi_j\rangle^{\otimes s})$ and $r_{ij,s}$ is a normalization constant, $i, j=1, 2, \cdots, m$; $s=1, 2, \cdots, M+2k$ and $M$ is an integer.

\label{thm9}
\end{thm}

The proof is shown in Appendix E.

\subsection{Pure states in a restricted set}

In this subsection, we generalize a previous overlap condition \cite{10} and obtain new superposing schemes of unknown states when multiple copies of the input states are available.

\begin{thm}
Let $\alpha, \beta$ be nonzero complex constants satisfying $|\alpha|^2+|\beta|^2=1$, $\rho_{X}$ be a known pure state in Hilbert space $\mathbb{H}^3$ with ${\rm dim}(\mathbb{H})\geq 2$, and $\rho_{\mu}$ be an unknown qubit state with vector representative  $|\mu\rangle=\alpha|0\rangle+\beta|1\rangle$.
\begin{itemize}
\item[(i)]
There exists a CP map ${\cal F}\in{\cal CP}(\mathbb{C}^2\otimes \mathbb{H}^2\otimes \mathbb{H}^2, \mathbb{H})$ for all the states $\rho_{\psi}, \rho_{\phi}$ on $\mathbb{H}$ satisfying the following inequalities
\begin{eqnarray}
&& {\rm tr}(\rho_X\otimes \rho_{\psi}\otimes
\rho_{\phi}\otimes \rho_{\psi})= c_1>0,
\nonumber
\\
&& {\rm tr}(\rho_X\otimes \rho_{\phi}
\otimes \rho_{\psi}\otimes \rho_{\phi})=c_2>0,
\label{eqn-61}
\end{eqnarray}
such that
\begin{eqnarray}
{\cal F}(\rho_{\mu}\rho_{\psi}^{\otimes 2}\rho_{\phi}^{\otimes 2})= \rho_{\varphi},
\label{eqn-63}
\end{eqnarray}
where the vector representative of $\rho_{\varphi}$ is defined by $|\varphi\rangle\propto\sqrt{r}(\alpha e^{i\theta_1}|\psi\rangle+
\beta e^{i\theta_2}|\phi\rangle)$ with $e^{i\theta_1}=\frac{\langle X|\phi\rangle|\psi\rangle|\phi\rangle}{|\langle X|\phi\rangle|\psi\rangle|\phi\rangle|}$ and $e^{i\theta_2}=\frac{\langle X|\psi\rangle|\phi\rangle|\psi\rangle}{|\langle X|\psi\rangle|\phi\rangle|\psi\rangle|}$ and $r$ is a normalization constant.
\item[(ii)]There exists a CP map ${\cal F}'\in{\cal CP}(\mathbb{C}^2\otimes \mathbb{H}^2\otimes \mathbb{H}^2, \mathbb{H})$ for all the states $\rho_{\psi}, \rho_{\phi}$ in $\mathbb{H}$ satisfying
\begin{eqnarray}
&&{\rm tr}(\rho_X\otimes \rho_{\psi}\otimes \rho_{\psi}
\otimes \rho_{\phi})= c'_1>0,
\nonumber
\\
&&{\rm tr}(\rho_X\otimes \rho_{\phi}\otimes \rho_{\phi}
 \otimes \rho_{\psi})= c'_2>0,
\label{eqn-63'}
\end{eqnarray}
such that
\begin{eqnarray}
{\cal F}'(\rho_{\mu} \rho_{\psi}^{\otimes 2}\rho_{\phi}^{\otimes 2})= \rho_{\varphi'},
\label{eqn-64'}
\end{eqnarray}
where the vector representative of $\rho_{\varphi'}$ is defined by $|\varphi'\rangle\propto\sqrt{r'}(\alpha e^{i\theta_3}|\psi\rangle+
\beta e^{i\theta_4} |\phi\rangle)$ with $e^{i\theta_3}=\frac{\langle X|\phi\rangle|\phi\rangle|\psi\rangle}{|\langle X|\phi\rangle|\phi\rangle|\psi\rangle|}$ and $e^{i\theta_4}=\frac{\langle X|\psi\rangle|\psi\rangle|\phi\rangle}{|\langle X|\psi\rangle|\psi\rangle|\phi\rangle|}$ and $r'$ is a normalization constant.
\end{itemize}

\label{thm10}
\end{thm}

\begin{figure}
\begin{center}
\resizebox{240pt}{165pt}{\includegraphics{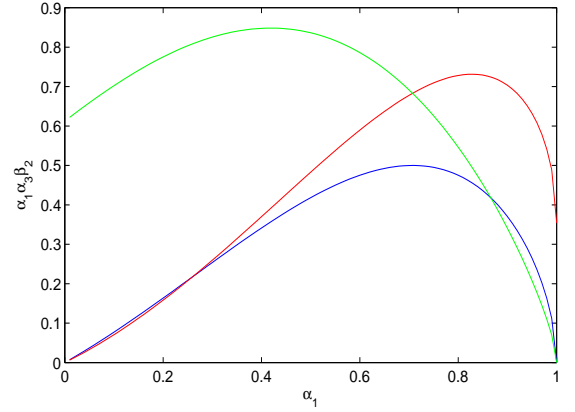}}
\end{center}
\caption{(Color online) Schematic figure of $c_1$ vias the $x$-axis component $\alpha_1$. Here, the vector $|X\rangle=|x_1\rangle|x_2\rangle|x_3\rangle$, we get that $\alpha_1\alpha_3\beta_2=c_1, \beta_1\beta_3\alpha_2=c_2$, where $\alpha_i=|\langle x_i |\psi\rangle|$ and $\beta_i=|\langle x_i |\phi\rangle|$, $i=1, 2, 3$. The blue line denotes the graph of $\{(\alpha_1,c_1)|\alpha_1\in[0,1]\}$ with $|x_1\rangle=|0\rangle, |x_2\rangle=\frac{1}{\sqrt{2}}(|0\rangle+|1\rangle)$ and $|x_3\rangle=|1\rangle$. The red line denotes the graph of $\{(\alpha_1,c_1)|\alpha_1\in[0,1]\}$ with $|x_1\rangle=|0\rangle, |x_2\rangle=\frac{1}{\sqrt{2}}(|0\rangle+|1\rangle)$ and $|x_3\rangle=\frac{1}{2}(|0\rangle+\sqrt{3}|1\rangle)$. The green line denotes the graph of $\{(\alpha_1,c_1)|\alpha_1\in[0,1]\}$ with $|x_1\rangle=|1\rangle, |x_2\rangle=\frac{1}{\sqrt{2}}(|0\rangle+|1\rangle)$ and $|x_3\rangle=\frac{1}{2}(|0\rangle+\sqrt{3}|1\rangle)$.}
\end{figure}

Before proving this theorem, we want to present comparisons of the condition in Eq.(\ref{eqn-61}) to the fixed overlap condition of Theorem 2 \cite{10}. It can be easily shown that the new condition in Eq.(\ref{eqn-61}) holds for the fixed overlap condition \cite{10}. However, the converse is not true. In particular, taking $|X\rangle=|x_1\rangle|x_2\rangle|x_3\rangle$ as an example, we obtain that $c_1=\alpha_1\alpha_3\beta_2$ and $c_2=\beta_1\beta_3\alpha_2$, where $\alpha_i=|\langle x_i |\psi\rangle|$ and $\beta_i=|\langle x_i |\phi\rangle|$, $i=1, 2, 3$. An evaluated example is shown in Fig.3. There exist superposed states that satisfy Eq.(\ref{eqn-61}) with the fixed state $\rho_X$ while they do not satisfy the condition of Theorem 2 \cite{10} with the fixed state $\rho_{x_1}$. It is because that there are two different values of $\alpha_1$ for some $c_1$.

{\it Proof of Theorem \ref{thm10}}. The proof procedure is similar to that of the superposing scheme \cite{10}. Define an auxiliary normalized vector $|v\rangle=\sqrt{c}(\sqrt{c_1}|0\rangle+\sqrt{c_2}|1\rangle)$, where $c$ is a normalization constant. Let ${\cal F}={\cal F}_5{\cal F}_4{\cal F}_3{\cal F}_2{\cal F}_1$, where
\begin{eqnarray}
{\cal F}_1: \rho&\mapsto &V_1\rho V^\dag_1,
\nonumber
\\
{\cal F}_2: \rho&\mapsto &V_2\rho V^\dag_2,
\nonumber
\\
{\cal F}_3: \rho&\mapsto &V_3\rho V^\dag_3,
\nonumber
\\
{\cal F}_4: \rho&\mapsto &V_4\rho V^\dag_4,
\nonumber
\\
{\cal F}_5: \rho&\mapsto&{\rm tr}_{1345}(\rho),
\end{eqnarray}
where
\begin{eqnarray}
V_{1}&=&|0\rangle\langle 0|\otimes I\otimes I\otimes I\otimes I+|1\rangle\langle 1|\otimes I\otimes I\otimes S_{4,5},
\nonumber
\\
V_{2}&=&|0\rangle\langle 0|\otimes I\otimes I\otimes I\otimes I+|1\rangle\langle 1|\otimes S_{2,3}\otimes I\otimes I,
\nonumber
\\
V_3&=&I\otimes I\otimes P_X,
\nonumber
\\
V_4&=&P_v\otimes I\otimes I \otimes I\otimes I.
\label{eqn46}
\end{eqnarray}
Here, $P_v=|v\rangle\langle v|$, $P_X=|X\rangle\langle X|$, $I$ denotes the identity operator, $S_{i,j}$ denotes the swapping operator of two pure states in the $i$-th and $j$-the subsystems, and ${\rm tr}_{1345}(\cdot{})$ is the partial trace over all the factors except the second. Its schematic circuit is shown in Fig.4, where one swapping operation $S_{3,4}$ is firstly used to obtain $\rho_\mu\rho_{\psi}\rho_{\phi}\rho_{\psi}\rho_{\phi}$ from $\rho_v\otimes \rho_{\psi}^{\otimes 2}\otimes \rho_{\phi}^{\otimes 2}$. ${\cal F}$ is completely positive and trace nonincreasing. The final state may be evaluated forward. The uniqueness is similar to that in Ref.\cite{10} except some different constructions, see Appendix E. The success probability is given by
\begin{eqnarray}
p_s&=&{\rm tr}[{\cal F}(\rho_\mu\otimes \rho_{\psi}^{\otimes 2}\otimes \rho_{\phi}^{\otimes 2})]
\nonumber
\\
&=&{\rm tr}[{\cal F}(\rho_\mu\rho_{\psi}\rho_{\phi}\rho_{\psi}\rho_{\phi})]
\nonumber
\\
&=&\frac{c_1c_2}{c_1+c_2}N^2_{\varphi}.
\end{eqnarray}
The map ${\cal F}$ cannot be rescaled to increase the success probability because of operator inequality $(V_4V_3V_2V_1)^\dag(V_4V_3V_2V_1)\leq I^{\otimes 5}$. If coefficients $\alpha$ and $\beta$ are fixed, one may construct a new CP map with a higher probability, see Appendix F.  $\hfill{}\Box$

\begin{figure}
\begin{center}
\resizebox{160pt}{90pt}{\includegraphics{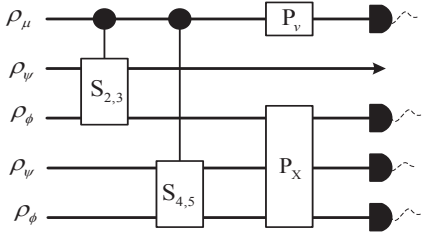}}
\end{center}
\caption{(Color online) Schematic circuit of CP mapping ${\cal F}$. Here, $P_X$ and $P_v$ denote the measurement induced by the operator $|X\rangle\langle X|$ and $|v\rangle\langle v|$, respectively. $\bullet$ denotes the controlling system. }
\end{figure}

It is possible to generalize the superposing map ${\cal F}$ for $k$ copies of the superposed states. The fixed state is defined in Hilbert space $\mathbb{H}^{2k-1}$. For convenience, we make use of the permutation group $\mathbb{P}_{2k-1}$ consisting of all the permutation operations on a finite set with $2k-1$ elements. For each permutation $g\in \mathbb{P}_{2k-1}$, there exists a general swapping operation $S_{g}$ induced by the permutation $g$ such that $S_g\in {\cal CP}(\mathbb{H}^{2k-1},\mathbb{H}^{2k-1})$. In particular, let $\mathbb{P}_{2k-1}$ be a permutation operated on the index set $\{1, 2, \cdots, 2k-1\}$. For each state $\rho_{\phi_{1,2,\cdots, 2k-1}}\in \mathbb{H}^{2k-1}$ (where subscripts $1, 2, \cdots, 2k-1$ denote the indexes of subsystems), then we define
\begin{eqnarray}
S_{g}: \rho_{\phi_{1, 2,\cdots, 2k-1}}\mapsto \rho_{\phi_{g(1,2,\cdots, 2k-1)}},
\end{eqnarray}
where $g(1, 2,\cdots, 2k-1)$ denotes the permutation of index pairs $(1, 2,\cdots, 2k-1)$ according to the permutation operator $g\in \mathbb{P}_{2k-1}$. Let
$\rho_{\Psi_{1,2,\cdots, 2k-1}}:=\rho_{\psi}^{\otimes k-1} \otimes \rho_{\phi}^{\otimes k}$ and $\rho_{\Phi_{1,2,\cdots, 2k-1}}:=\rho_{\psi}^{\otimes k} \otimes \rho_{\phi}^{\otimes k-1}$.

\begin{thm}
Let $\alpha, \beta$ be nonzero complex constants satisfying $|\alpha|^2+|\beta|^2=1$, $\rho_X$ be a known pure state in Hilbert space $\mathbb{H}^{2k-1}$ with ${\rm dim}(\mathbb{H})\geq 2$, and $\rho_{\mu}$ be an unknown qubit state on $\mathbb{C}^2$ with vector representative $|\mu\rangle\propto\alpha|0\rangle+\beta|1\rangle$. For $\tau_1, {\tau}_2\in \mathbb{P}_{2k-1}$. Then there exists a CP map ${\cal F}_{\tau_1,\tau_2}\in{\cal CP}(\mathbb{C}^2\otimes \mathbb{H}^k\otimes \mathbb{H}^k, \mathbb{H})$ for all pure states $\rho_{\psi}^{\otimes k}, \rho_{\phi}^{\otimes k}$ in $\mathbb{H}^k$ satisfying
\begin{eqnarray}
&&{\rm tr}(\rho_X\otimes \rho_{\Psi_{\tau_1(1,2,\cdots, 2k-1)}})= c_1>0,
\nonumber
\\
&&
{\rm tr}(\rho_X\otimes \rho_{\Phi_{{\tau}_2(1,2,\cdots, 2k-1)}})= c_2>0,
\label{eqn-67}
\end{eqnarray}
such that
\begin{eqnarray}
{\cal F}_{\tau_1,\tau_2}(\rho_\mu \rho_{\psi}^{\otimes k}\rho_{\phi}^{\otimes k})=\rho_{\varphi},
\label{eqn-68}
\end{eqnarray}
where the vector representative of $\rho_{\varphi}$ is defined by $|\varphi\rangle\propto\sqrt{r}(\alpha e^{i\theta_1}|\psi\rangle+
\beta e^{i\theta_2}|\phi\rangle)$ with $e^{i\theta_1}=\frac{\langle X|\Phi_{{\tau}_2(1,2,\cdots, 2k-1)}\rangle}{|\langle X|\Phi_{{\tau}_2(1,2,\cdots, 2k-1)}\rangle|}$ and $e^{i\theta_2}=\frac{\langle X|\Psi_{\tau_1(1,2,\cdots, 2k-1)}\rangle}{|\langle X|\Psi_{\tau_1(1,2,\cdots, 2k-1)}\rangle|}$ and $r$ is a normalization constant.

\label{thm10g}
\end{thm}

In Theorem \ref{thm10g}, $|\Psi_{\tau_i(1,2,\cdots, 2k-1)}\rangle$ is the vector representative of the pure state $\rho_{\Psi_{{\tau}_i(1,2,\cdots, 2k-1)}}$, and $|\Phi_{\tau_i(1,2,\cdots, 2k-1)}\rangle$ is the vector representative of the pure state $\rho_{\Phi_{{\tau}_i(1,2,\cdots, 2k-1)}}$. The proof of this theorem is similar to that of Theorem \ref{thm10g}, see Appendix G.

In general, we can superpose $n$ independent states as follows. Let
$\rho_{\Psi^{j}_{1,2,\cdots, nk-1}}:=\otimes_{t=1}^n\rho_{\psi_t}^{\otimes k_{jt}}$, where $k_{jt}$ is an integer satisfying $k_{jt}=k$ for $j\not=t$ and $k_{jt}=k-1$ for $j=t$.

\begin{cor}
Let $\rho_X$ be a known pure state on Hilbert space $\mathbb{H}^{nk-1}$ with ${\rm dim}(\mathbb{H})\geq 2$, and $\rho_{\mu}$ be an unknown state in $\mathbb{C}^n$ with vector representative $|\mu\rangle=\sum_{j=0}^{n-1}\alpha_{j}|j\rangle$. For any permutations $\tau_1, \tau_2, \cdots, \tau_{n}\in\mathbb{P}_{nk-1}$, there exists a CP map ${\cal F}_{\tau_1, \tau_2, \cdots, \tau_{n}}\in{\cal CP}(\mathbb{C}^n\otimes \mathbb{H}^{nk}, \mathbb{H}^k)$ for all pure states $\rho_{\psi_1}^{\otimes k}, \rho_{\psi_2}^{\otimes k}, \cdots, \rho_{\psi_n}^{\otimes k}$ in $\mathbb{H}^k$ satisfying
\begin{eqnarray}
{\rm tr}(\rho_X\otimes \rho_{\Psi^i_{\tau_i(1,2,\cdots, nk-1)}})= c_i>0, i=1, 2, \cdots, n
\label{eqn-70}
\end{eqnarray}
such that
\begin{eqnarray}
{\cal F}_{\tau_1, \tau_2, \cdots, \tau_{n}}(\rho_\mu\otimes (\otimes_{i=1}^n \rho_{\psi_i}^{\otimes k}))=\rho_{\varphi},
\label{eqn-71}
\end{eqnarray}
where the vector representative of $\rho_{\varphi}$ is defined by $|\varphi\rangle\propto\sqrt{r}\sum_{j=0}^{n-1}\alpha_j e^{i\theta_j}|\psi_j\rangle$ with $e^{i\theta_j}=\frac{\langle X|\varphi_{\tau_j(1,2,\cdots,nk-1)}\rangle}
{|\langle X|\varphi_{\tau_j(1,2,\cdots,nk-1)}\rangle|}$ and $r$ is a normalization constant.

\end{cor}

In this corollary, $|\varphi_{\tau_j(1, 2,\cdots, nk-1)}\rangle$ is the vector representative of the pure state $\rho_{\varphi_{\tau_j(1,2,\cdots,nk-1)}}$, $j=1, 2, \cdots, n$. The proof is similar to that of Theorem \ref{thm10}, see Appendix H.

\section{Conclusions}

We have firstly presented one theorem that may extend the no-superposing theorem to forbid superposing of an unknown state and a fixed state. This theorem has been generalized for multiple copies of the input states. The new no-go theorems include the no-cloning theorem, no-deleting theorem and no-superposing theorem as special cases. And then, we presented unified schemes of imperfect and perfect transforming (cloning and deleting) of the pure states in a finite set. Similar to the unambiguous state discriminations, all the schemes are completely characterized with the linear independence of the input states. Finally, new schemes are constructed to superpose unknown states. These superposing schemes are completed when some permutations of input states have fixed overlaps with a fixed state.

{\bf Acknowledgements}

We thank the comments of reviewers. We thank the helps of Luming Duan and M. Orgun. This work was supported by the National Natural Science Foundation of China (No. 61303039), Sichuan Youth Science and Technique Foundation (No.2017JQ0048), Fundamental Research Funds for the Central Universities (Nos.2682014CX095, XDJK2016C043), Chuying Fellowship, CSC Scholarship, and EU ICT COST CryptoAction (No.IC1306).

\section*{Appendix A. Completing Proof of the Theorem 1}

Assume that there exists a unitary map ${\cal U}$ with the matrix representative $U$ satisfying Eq.(3) for all the unknown states $\rho_{h}$ on $\mathbb{H}$. For two basis states $\rho_0$ and $\rho_1$, it follows that
\begin{eqnarray}
U|0\rangle|\Sigma\rangle|P_0\rangle&\propto&
\sqrt{p_0}|0\rangle|\Sigma_0\rangle|P_1\rangle
+\sqrt{p_1}|{{\Phi}_0}\rangle,
\label{app1'}
\\
U|1\rangle|\Sigma\rangle|P_0\rangle&\propto&
\sqrt{p'_0}|{\chi'}\rangle|{\Sigma_1}\rangle|{P_1}\rangle
+\sqrt{p'_1}{{\Phi}_1}\rangle,
\label{app1''}
\end{eqnarray}
where the vector representative of $\rho_{\chi'}$ is defined by $|\chi'\rangle\propto\alpha|1\rangle+\beta|0\rangle$. $\rho_{{\Sigma}_0}$ and $\rho_{{\Sigma}_1}$ are the output states of the ancillary system, and generally depend on the input states.

For each pure state $\rho_{{h}}$ with the vector representative $|h\rangle\propto a|0\rangle+b|1\rangle$ satisfying $|a|^2+|b|^2=1$, we obtain
\begin{eqnarray}
U|h\rangle|{\Sigma}\rangle|P_0\rangle
\propto
\sqrt{p''_0}|\varphi_1\rangle|\Sigma_2\rangle|P_1\rangle
+\sqrt{p''_1}|{\Phi}_2\rangle
\label{app2}
\end{eqnarray}
where the vector representative of $\rho_{\varphi_1}$ is defined by $|\varphi_1\rangle\propto\sqrt{r''}({\alpha}|h\rangle+{\beta}|0\rangle)$ and $r''$ is a normalization constant, and $\rho_{{\Sigma}_2}$ is the output state of the ancillary system. Moreover, from the linearity of quantum operations and Eqs.(\ref{app1'}) and (\ref{app1''}), we obtain
\begin{eqnarray}
U|\varphi\rangle|\Sigma\rangle|P_0\rangle
&\propto&|\Psi\rangle|P_1\rangle+a\sqrt{p_1}|{\Phi}_0\rangle
\nonumber
\\
&&+b\sqrt{p'_1}|{\Phi}_1\rangle,
\label{app3}
\end{eqnarray}
where the vector $|\Psi\rangle=a\sqrt{p_0}|0\rangle|{\Sigma}_0\rangle
+b\sqrt{p'_0}|\chi'\rangle|\Sigma_1\rangle$. Let $|{\Sigma}_i\rangle=a_i|0\rangle+b_i|1\rangle$ with $|a_i|^2+|b_i|^2=1$, $i=1, 2, 3$. From Eqs.(\ref{app2}) and (\ref{app3}), we get $a=0$, which is a contradiction to the assumption of $\rho_h$.

\section*{Appendix B. Proof of Theorem 3}

Similar to the proof of Theorem 2, we only need to prove the result for $0<|\beta|<1$. Assume that there exists a unitary map ${\cal U}$ with the  matrix representative $U$ for all the states $\rho_{\psi}$ in $\mathbb{H}$ such that
\begin{eqnarray}
U|\psi\rangle^{\otimes k}|\Sigma\rangle|P_0\rangle
&\propto&
\sqrt{p_0}|\varphi\rangle^{\otimes n}|0\rangle^{\otimes N+k-n}|P_1\rangle
\nonumber
\\
&& +\sqrt{p_1}|\Phi\rangle,
\label{app9}
\end{eqnarray}
where the vector representative of $\rho_{\varphi}$ is defined by $|\varphi\rangle\propto\sqrt{r}(\alpha|\psi\rangle+\beta|\phi\rangle)$ and $r$ is a normalization constant. $|\Sigma\rangle$ is the vector representative of an ancillary state $\rho_{\Sigma}$ on Hilbert space $\mathbb{H}_A=\mathbb{H}^N$ with $N>n$. $|P_0\rangle$ and $|P_1\rangle$ are the respective vector representative of orthogonal states $\rho_{P_0}$ and $\rho_{P_1}$ on probe space $\mathbb{H}_P$ with ${\rm dim}(\mathbb{H}_P)\gg 2$. Let $\rho_{e^{i\theta}\varphi}$ be a new state. From Eq.(\ref{app9}) it follows that
\begin{eqnarray}
U(e^{i\theta}|\psi\rangle^{\otimes k}|\Sigma\rangle|P_0\rangle)
&\propto& \sqrt{p_0} |\varphi'\rangle^{\otimes n}|0\rangle^{\otimes N+k-n}|P_0\rangle
\nonumber
\\
&&+\sqrt{p_1}|\Phi'\rangle
\label{app10}
\end{eqnarray}
where the vector representative of $\rho_{\varphi'}$ is defined by $|\varphi'\rangle\propto\sqrt{r'}(\sqrt{\alpha}e^{i\theta}|\varphi\rangle
+\sqrt{\beta}|\phi\rangle)$ and $r'$ is a normalization constant. Note that $|\psi\rangle\propto e^{i\theta}|\psi\rangle$, i.e., the pure state $\rho_{e^{i\theta}\psi}$ is physically undiscriminating from the pure state $\rho_{\psi}$. From Eq.(\ref{app9}) it follows that
\begin{eqnarray}
U(e^{i\theta}|\psi\rangle^{\otimes k}|\Sigma\rangle|P_0\rangle)\propto
U|\psi\rangle^{\otimes k}|\Sigma\rangle|P_0\rangle.
\label{app10'}
\end{eqnarray}
Eqs. (\ref{app10}) and (\ref{app10'}) lead to $\beta=0$, which is a contradiction to the assumption of $\beta$.

\section*{Appendix C. The proof of inequality (\ref{eqn-21})}

From Eq.(\ref{eqn-13}) and the triangle inequality, we obtain
\begin{eqnarray}
|a_{ij}|^k&\leq&
\sum_{s=1}^{M+k}\sqrt{p_{is}p_{js}}(\sqrt{r_{is}r_{js}}
|\alpha_{is}\alpha_{js}|\times|a_{ij}|^s
\nonumber
\\
&&
+|\alpha_{is}\beta_{js}\lambda_{ij,s}|
+|\beta_{is}\alpha_{js}\lambda_{ji,s}|
+|\beta_{is}\beta_{js}|)
\nonumber
\\
&&
+\sum_{\ell=M+k+1}^N\sqrt{q_{i\ell}q_{j\ell}}.
\label{eqn-18}
\end{eqnarray}
Note that for all the complex constants $\alpha_{is}, \beta_{js}$ with $|\alpha_{is}|, |\beta_{js}|\leq 1$, we have $|\alpha_{is}\beta_{js} \lambda_{ij,s}|+|\beta_{is}\alpha_{js}\lambda_{ji,s}|\leq \max\{|\lambda_{ij,s}|,|\lambda_{ji,s}|\} \times
(|\alpha_{is}\beta_{js}|+|\beta_{is}\alpha_{js}|)$, and
$|\alpha_{is}\beta_{js}|+|\beta_{is}\alpha_{js}| \leq\frac{1}{2}( |\alpha_{is}|^2+|\beta_{js}|^2 +|\beta_{is}|^2+|\alpha_{js}|^2)=1$ from the equalities $|\alpha_{is}|^2+|\beta_{is}|^2=|\alpha_{js}|^2+|\beta_{js}|^2=1$. Hence, from the arithmetic-geometric average inequality, Eq.(\ref{eqn-18}) reduces to
\begin{eqnarray}
|a_{ij}|^k&\leq&
\sum_{s=1}^{M+k}{p}_{ij,s}(\max\{|\lambda_{ij,s}|,|\lambda_{ji,s}|\}
\nonumber
\\
&&+\alpha_{ij,s}{r}_{ij,s}|a_{ij}|^s+\beta_{ij,s}-1)+1,
\label{eqn-19}
\end{eqnarray}
where ${p}_{ij,s}=(p_{is}+p_{js})/2$, $\alpha_{ij,s}=|\alpha_{is}\alpha_{js}|$, $\beta_{ij,s}=|\beta_{is}\beta_{js}|$ and ${r}_{ij,s}=\sqrt{r_{is}r_{js}}$, and we have taken use of the equalities $\sum_{s=1}^{M+k}p_{is}+\sum_{\ell=M+k+1}^N q_{i\ell}=1$, $i=1, 2, \cdots, m$. Denote $D_{ij,t}:=2(1-|a_{ij}|^t)$ with $t=k$ or $s$. Since $1-\beta_{ij,s} \geq \alpha_{ij,s}$ for $|\alpha_{is}|^2+|\beta_{is}|^2=|\alpha_{js}|^2+|\beta_{js}|^2=1$, Eq.(\ref{eqn-19}) yields to inequality (\ref{eqn-21}).

\section*{Appendix D. The proof of inequality (\ref{eqn-37'})}

From Eq.(\ref{eqn-55}) we obtain
\begin{eqnarray}
|a_{i_1j_1}a_{i_2j_2}|
&\leq& \sqrt{p_{i_1i_2}p_{j_1j_2}}
\sqrt{r_{i_1i_2}r_{j_1j_2}}
\nonumber
\\
&&
\times(|\alpha_{i_1}\alpha_{j_1}a_{i_1j_1}|
+|\beta_{i_2}\beta_{j_2}a_{i_2j_2}|
\nonumber
\\
&&
+|\alpha_{i_1}\beta_{j_2}a_{i_1j_2}|
+|\beta_{i_2}\alpha_{j_1}a_{i_2j_1}|
)
\nonumber
\\
&&
+\sqrt{q_{i_ii_2}q_{j_1j_2}}
\nonumber
\\
&\leq&\frac{1}{3}\sqrt{p_{i_1i_2}p_{j_1j_2}}
\sqrt{r_{i_1i_2}r_{j_1j_2}}
\nonumber
\\
&&
\times (2|\alpha_{i_1}|^2
+2|\alpha_{j_1}|^2
+2|\beta_{j_2}|^2
+2|\beta_{i_2}|^2
\nonumber
\\
&&
+|a_{i_1j_1}|+|a_{i_2j_2}|
+|a_{i_2j_1}|+|a_{i_1j_2}|)
\nonumber
\\
&&
+\sqrt{q_{i_ii_2}q_{j_1j_2}}
\nonumber\\
&\leq &\frac{1}{3}\sqrt{p_{i_1i_2}p_{j_1j_2}}
\sqrt{r_{i_1i_2}r_{j_1j_2}}
\nonumber
\\
&&
\times(|a_{i_1j_1}|+|a_{i_2j_2}|
+|a_{i_2j_1}|
\nonumber
\\
&&
+|a_{i_1j_2}|+4)+\sqrt{q_{i_ii_2}q_{j_1j_2}},
\label{eqn-37}
\end{eqnarray}
where the second inequality is derived from the inequality
$x^2+y^2+z\geq x^3+y^3+z^3\geq 3 xyz$ for $0\leq x,y,z\leq 1$, and the last equality is derived from the equalities $|\alpha_{i}|^2+|\beta_{i}|^2=1$, $i=1, 2, \cdots, m$. By using the arithmetic inequality for ${p}_{i_ii_2}{p}_{j_1j_2}$ and $q_{i_ii_2}q_{j_1j_2}$, Eq.(\ref{eqn-37}) yields to inequality (\ref{eqn-37'}).

\section*{Appendix E. Proof of Theorem 7}
Note that the vector representatives $|\psi_1\rangle^{\otimes k}|\psi_1\rangle^{\otimes k}$, $|\psi_1\rangle^{\otimes k}|\psi_2\rangle^{\otimes k}$, $\cdots, |\psi_m\rangle^{\otimes k}|\psi_m\rangle^{\otimes k}$ are linearly dependent if and only if the vectors $|\psi_1\rangle^{\otimes k}, |\psi_2\rangle^{\otimes k}, \cdots, |\psi_m\rangle^{\otimes k}$ are linearly dependent. Similar to the proof of Theorem 5, the necessity is easily followed from the linearity of quantum operations and the superposition principle. Now, we prove that these CP maps will be presented in a unified form with a nontrivial probability in the following. In fact, from Stinespring's Theorem \cite{Stin}, assume that there exists a unitary map with matrix representative $U$ such that
\begin{eqnarray}
&&U|\psi_i\rangle^{\otimes k}|{\psi_j}\rangle^{\otimes k}|{\Sigma}\rangle|{P_0}\rangle
\nonumber
\\
&\propto&\sum_{s=1}^{M+2k}\sqrt{p_{ij,s}}|{\varphi_{ij,s}}\rangle|0\rangle
^{M+2k-s}|P_s\rangle
\nonumber
\\
&&+\sum^N_{\ell=M+2k+1}\sqrt{q_{ij,\ell}}|{\Phi}_{ij,\ell}\rangle|P_\ell\rangle
\end{eqnarray}
Here, $|\Sigma\rangle$ is the vector representative of an ancillary state $\rho_{\Sigma}$ on Hilbert space $\mathbb{H}_A=\mathbb{H}^M$ with $M>\max\{k,s\}$. $p_{ij,s}$ is the success probability of producing $\rho_{\varphi_{ij,s}}$. $|P_0\rangle, |P_1\rangle, \cdots, |P_N\rangle$ are the vector representatives of orthogonal states $\rho_{P_0}, \rho_{P_1}, \cdots, \rho_{P_N}$ on probe space $\mathbb{H}_P$ with ${\rm dim}(\mathbb{H}_P)>N+1$. $|{\Phi}_{ij,\ell}\rangle$ is the vector representative of a normalized failure state $\rho_{{\Phi}_{ij,\ell}}$ on $\mathbb{H}^{k}\otimes \mathbb{H}^{M}$ with the probability $q_{ij,\ell}$ which satisfies $\sum_{s=1}^{M+2k}p_{ij,s}+\sum_{\ell=M+2k+1}^Nq_{ij,\ell}=1$, $i,j=1, 2, \cdots, m$.  Each CP map ${\cal F}_j$ is defined by ${\cal F}_j=|0\rangle^{M+k-j}\langle0|{}^{M+k-j}\circ |P_j\rangle\langle P_j|\circ {\cal U}$, $j=1, 2, \cdots, M+k$.

For the input states $\rho_{\psi_{i_1}}^{\otimes k}\rho_{\psi_{i_2}}^{\otimes k}$ and $\rho_{\psi_{j_1}}^{\otimes k}\rho_{\psi_{j_2}}^{\otimes k}$, taking the inner product of the vector representatives of the output states, it follows that
\begin{eqnarray}
a_{i_1j_1}^{k}a_{i_2j_2}^{k}
&=&\sum_{s=1}^{M+2k}\sqrt{p_{i_1i_2,s}p_{j_1j_2,s}
r_{i_1i_2,s}r_{j_1j_2,s}}h^{(s)}_{i_1i_2,j_2j_2}
\nonumber
\\
&&+\sum_{\ell=M+2k+1}^N
\sqrt{q_{i_1i_2,\ell}q_{j_1j_2,\ell}}b_{i_1i_2,j_2j_2,\ell},
\label{app13}
\end{eqnarray}
where $a_{ij}=\langle\psi_i|\psi_j\rangle$, $b_{i_1i_2,j_2j_2,\ell}=\langle \Phi_{i_1i_2,\ell}|\Phi_{j_1j_2,\ell}\rangle$, and
$h^{(s)}_{i_1i_2,j_2j_2}=\alpha^*_{i_1i_2}\alpha_{j_1j_2}a^s_{i_1j_1}
+\beta^*_{i_1i_2}\beta_{j_1j_2}a^s_{i_2j_2}
+\alpha^*_{i_1i_2}\beta_{j_1j_2}a^s_{i_1j_2}
+\beta^*_{i_1i_2}\alpha_{j_1j_2}a^s_{i_2j_1}$ for all integers $1\leq i, j, i_1, i_2, j_1, j_2\leq m$ and $M+2k+1\leq \ell\leq N$. Eq.(\ref{app13}) may be briefly represented by a matrix equation:
\begin{eqnarray}
A^{\circ k}=\sum_{s=1}^{M+2k}\Lambda_s B_s \Lambda_s^\dag+
\sum_{\ell=M+2k+1}^{N}Q_\ell,
\label{app14}
\end{eqnarray}
where
\begin{eqnarray*}
&&A=\left[a_{i_1j_1}a^*_{i_2j_2}\right]_{m^2\times m^2},
\\
&&\Lambda_s={\rm diag}(p_{11,s}, p_{12,s}, \cdots, p_{mm,s})_{m^2\times m^2},
\\
&&H_s=
\left[
\sqrt{r_{i_1i_2,s}r_{j_1j_2,s}}h^{(s)}_{i_1i_2,j_2j_2}
\right]_{m^2\times m^2}
\\
&&Q_\ell=\left[
\sqrt{q_{i_1i_2,\ell}q_{j_1j_2,\ell}}b_{i_1i_2,j_2j_2,\ell}\right]_{m^2\times m^2},
\end{eqnarray*}
Here, the rows or columns of these matrices are represented by two-bit series $i_1i_2$ or $j_1j_2$, respectively. It is sufficient to prove Eq.(\ref{app14}) with physical realizable matrices $Q_\ell$, $\ell=M+2k+1, \cdots, L$. In fact, $A^{\circ k}$ is positive definite from Theorem 4 because $A$ is positive definite from the linear independence of the vector representatives $|\psi_1\rangle^{\otimes k}, |\psi_2\rangle^{\otimes k}, \cdots, |\psi_m\rangle^{\otimes k}$. The matrix $\sum_{s=1}^{M+2k}\Lambda_s H_s \Lambda_s^\dag$ is Hermite. From Lemma 1, the matrix $A^{\circ k}-\sum_{s=1}^{M+2k}\Lambda_s B_s \Lambda_s^\dag$ is positive definite when matrices $\Lambda_1, \Lambda_2, \cdots, \Lambda_{M+2k}$ satisfy the following inequality
\begin{eqnarray}
\|\sum_{s=1}^{M+2k}\Lambda_s B_s \Lambda_s^\dag\|_2<\|(A^{\circ k})^{-1}\|_2^{-1}
\label{app17}
\end{eqnarray}
Therefore, there exists a unitary matrix $V$ such that
\begin{eqnarray}
V(A^{\circ k}-\sum_{s=1}^{M+2k}\Lambda_s B_s \Lambda_s^\dag)V^\dag={\rm diag}(\lambda_{1}, \lambda_{2}, \cdots, \lambda_{m^2})
\label{app18}
\end{eqnarray}
where $\lambda_{1}, \lambda_{2}, \cdots, \lambda_{m^2}$ are all the eigenvalues of $A^{\circ k}-\sum_{s=1}^{k_3}\Lambda_s H_s \Lambda_s^\dag$, and satisfy $\lambda_j>0, j=1, 2, \cdots, m^2$. Define $Q_\ell=V^\dag {\rm diag}(\lambda_{1,\ell}$, $\lambda_{2,\ell}, \cdots, \lambda_{m^2,\ell})V$, where $\lambda_{1,\ell}$, $\lambda_{2,\ell}, \cdots, \lambda_{m^2,\ell}$ are positive constants and satisfy $\sum_{\ell=M+2k+1}^NQ_\ell={\rm diag}(\lambda_{1}, \lambda_{2}, \cdots, \lambda_{m^2})$. So, $Q_\ell$ is positive definite and then physically realizable, $\ell=M+2k+1, M+2k+2, \cdots, N$. This completes the proof.  $\hfill{} \Box$

In the following, the bound of the success probability is proved in terms of the state metric \cite{14}. In detail, from Eq.(\ref{app13}) we obtain
\begin{widetext}
\begin{eqnarray}
|a_{i_1j_1}a_{i_2j_2}|^k
&\leq& \sum_{s=1}^{M+2k}\sqrt{p_{i_1i_2,s}p_{j_1j_2,s}}
\sqrt{r_{i_1i_2,s}r_{j_1j_2,s}}
(|\alpha_{i_1i_2}\alpha_{j_1j_2}|\cdot|a_{i_1j_1}|^s
+|\beta_{i_1i_2}\beta_{j_1j_2}|\cdot|a_{i_2j_2}|^s
\nonumber
\\
&&
+|\beta_{i_1i_2}\alpha_{j_1j_2}|\cdot|a_{i_2j_1}|^s
+|\alpha_{i_1i_2}\beta_{j_1j_2}|\cdot|a_{i_1j_2}|^s)
+\sum_{\ell=M+2k+1}^N\sqrt{q_{i_1i_2,\ell}q_{j_1j_2,\ell}}
\nonumber
\\
&\leq&\frac{1}{3}\sum_{s=1}^{M+2k}\sqrt{p_{i_1i_2,s}p_{j_1j_2,s}}
\sqrt{r_{i_1i_2,s}r_{j_1j_2,s}}
[2|\alpha_{i_1i_2}|^2+2|\alpha_{j_1j_2}|^2
+2|\beta_{i_1i_2}|^2
\nonumber
\\
&&
+2|\beta_{j_1j_2}|^2
+|a_{i_1j_1}|^s+|a_{i_1j_2}|^s+|a_{i_2j_1}|^s+|a_{i_2j_2}|^s]
+\sum_{\ell=M+2k+1}^N\sqrt{q_{i_1i_2,\ell}q_{j_1j_2,\ell}}
\nonumber\\
&=&\frac{1}{3}\sum_{s=1}^{M+2k}\sqrt{p_{i_1i_2}p_{j_1j_2}}
\sqrt{r_{i_1i_2}r_{j_1j_2}}
(|a_{i_1j_1}|^s+|a_{i_2j_2}|^s+|a_{i_2j_1}|^s+|a_{i_1j_2}|^s+4)
\nonumber\\
&&
+\sum_{\ell=M+2k+1}^N\sqrt{q_{i_1i_2,\ell}q_{j_1j_2,\ell}},
\label{app19}
\end{eqnarray}
\end{widetext}
where the second inequality is derived from the inequality
$x^2+y^2+z\geq x^3+y^3+z^3\geq 3 xyz$ for $0\leq x,y,z\leq 1$, and the last equality is derived from the equalities $|\alpha_{ij}|^2+|\beta_{ij}|^2=1$ for any integers $1\leq i, j\leq m$ and $1\leq s\leq M+2k$.

By using the arithmetic inequality for $p_{i_1i_2,s}p_{j_1j_2,s}$ and $q_{i_1i_2,s}q_{j_1j_2,s}$, and the equality $\sum_{s=1}^{M+2k}p_{ij,s}+\sum_{\ell=M+2k+1}^Nq_{ij,\ell}=1$ for all $i,j=1, 2, \cdots, m$, Eq.(\ref{app19}) leads to
\begin{eqnarray}
D_{i_1j_1,k}D_{i_2j_2,k}
&\leq&
\frac{2}{3}\sum_{s=1}^{M+2k}p_{i_1i_2,j_1j_2,s}
[r_{i_1i_2,j_1j_2}(16-D_{i_1j_1,s}
\nonumber
\\
& &
-D_{i_1j_2,s}-D_{i_2j_2,s}
-D_{i_2j_1,s})-6]
\nonumber
\\
&&
+4D_{i_1j_1,k}+4D_{i_2j_2,k},
\label{app20}
\end{eqnarray}
where $r_{i_1i_2,j_1j_2}=\sqrt{r_{i_1i_2,s}r_{j_1j_2,s}}$, $p_{i_1i_2,j_1j_2,s}=(p_{i_1i_2,s}+p_{j_1j_2,s})/2$ and $D_{ij,s}=2(1-|a_{ij}|^s)$. The inequality has generalized the bound in Eq.(\ref{eqn-37'}).

\section*{Appendix F. Proof of the uniqueness of Theorem 8}

In this appendix, we complete the proof of Theorem 8 according to the proof \cite{10}. Let $|v\rangle=\sqrt{c}(\sqrt{c_1}|0\rangle+\sqrt{c_2}|1\rangle)$ be an ancillary vector and $\rho_{\psi}$ and $\rho_{\phi}$ be states on  $\mathbb{H}$, $\rho_X\in \mathbb{H}^{3}$ satisfy the conditions
\begin{eqnarray}
{\rm tr}(\rho_X\rho_{\psi}\rho_{\phi}\rho_{\psi})=c_1>0,
{\rm tr}(\rho_X\rho_{\phi}\rho_{\psi}\rho_{\phi})= c_2>0.
\label{app21}
\end{eqnarray}
Let ${\cal F}\in {\cal CP}(\mathbb{C}^2\otimes \mathbb{H}^{2}\otimes \mathbb{H}^{2},\mathbb{H})$ be a CP map satisfying
\begin{eqnarray}
{\cal F}(\rho_\mu \rho_{\psi}^{\otimes 2}
 \rho_{\phi}^{\otimes 2})=\rho_{\varphi},
\label{app22}
\end{eqnarray}
where the vector representative $|\varphi\rangle$ is given by
\begin{eqnarray*}
|\varphi\rangle\propto
\sqrt{r}(\alpha e^{i\theta_1}
|\psi\rangle+
\beta e^{i\theta_2}|\phi\rangle)
\end{eqnarray*}
and $r$ is a normalization constant dependent of $\alpha, \beta, |\psi\rangle$ and $|\phi\rangle$, $e^{i\theta_1}=\frac{\langle X|\phi\rangle|\psi\rangle|\phi\rangle}{|\langle X|\phi\rangle|\psi\rangle|\phi\rangle|}$ and $e^{i\theta_2}=\frac{\langle X|\psi\rangle|\phi\rangle|\psi\rangle}{|\langle X|\psi\rangle|\phi\rangle|\psi\rangle|}$. Let $\{F_i|F_i:\mathbb{C}^2\otimes \mathbb{H}^{4}\to \mathbb{H}\}_{i\in J}$, form the Kraus decomposition of ${\cal F}$. Using the analogous procedure to Theorem 1 \cite{10}, we get
\begin{eqnarray}
F_i(\rho_\mu\rho_{\psi}^{\otimes 2}\rho_{\phi}^{\otimes 2})F^\dag_i
= \rho_{\varphi}, \mbox{ for all } i\in J
\label{app23}
\end{eqnarray}
up to a global normalization factor. Now consider the single Kraus operator $F_i$. It follows that
\begin{eqnarray}
F_i|\mu\rangle|\psi\rangle^{\otimes 2}|\phi\rangle^{\otimes 2}
=a(\alpha e^{i\theta_1}|\psi\rangle+\beta e^{i\theta_2}|\phi\rangle)
\label{app24}
\end{eqnarray}
for all vectors $|\psi\rangle$, $|\phi\rangle\in \mathbb{H}$ satisfying the condition in Eq.(\ref{app21}) and a constant $a$ which is dependent of $\alpha, \beta, |\psi\rangle$ and $|\phi\rangle$. This definition is unique up to a global factor. In detail, from the linearity of the left side of Eq. (\ref{app24}), $a$ is independent of $\alpha$ and $\beta$. Moreover, from the linearity of $F_i$ and Eq.(\ref{app24}), it follows that
\begin{eqnarray}
a(|\psi\rangle^{\otimes 2},|\phi\rangle^{\otimes 2})
=a(e^{i\theta'_1}|\psi\rangle^{\otimes 2},e^{i\theta'_2}|\phi\rangle^{\otimes 2})
\label{app25}
\end{eqnarray}
for arbitrary phases $\theta'_i$. Hence, we can assume the following form
\begin{eqnarray}
&|\phi\rangle|\psi\rangle|\phi\rangle=\sqrt{c_1}|X\rangle + \sqrt{d_1}|\Phi^{\bot}\rangle,
\label{app26}
\\
&|\psi\rangle|\phi\rangle|\psi\rangle=\sqrt{c_2}|X\rangle + \sqrt{d_2}|\Psi^{\bot}\rangle,
\label{app27}
\end{eqnarray}
where $|X\rangle$ is the vector representative of the state $\rho_X$ and  $c_i+d_i=1$, $|\Phi^{\bot}\rangle$ and $|\Psi^{\bot}\rangle$ are normalized orthogonal complements of $|X\rangle$. From Eq.(\ref{app24}) we obtain
\begin{eqnarray}
&&F_i\otimes I\otimes I[|\mu\rangle(\sqrt{c_1}|\tilde{X}\rangle+d_1|\tilde{\Phi}^{\bot}\rangle)
(\sqrt{c_2}|\tilde{X}\rangle+d_2|\tilde{\Psi}^{\bot}\rangle)]
\nonumber
\\
&=&\tilde{a}((\alpha\sqrt{c_1}+\beta\sqrt{c_2})|\tilde{X}\rangle
+\alpha\sqrt{d_1}|\Psi^{\bot}\rangle
\nonumber
\\
&&
+\beta \sqrt{d_2}|\Phi^{\bot}\rangle)|\phi\rangle|\psi\rangle,
\label{app28}
\end{eqnarray}
where the vectors $|\tilde{X}\rangle=(S_{1,2}\otimes I)|X\rangle$, $|\tilde{\Phi}^{\bot}\rangle=(S_{1,2}\otimes I)|{\Phi}^{\bot}\rangle$ and $|\tilde{\Psi}^{\bot}\rangle=(S_{1,2}\otimes I)|{\Psi}^{\bot}\rangle$. For normalized vectors $|\tilde{\Psi}^{\bot}\rangle$ and $|\tilde{\Phi}^{\bot}\rangle$, the function
\begin{eqnarray}
(\theta_1, \theta_2)\mapsto \tilde{a}(e^{i\theta_1}|\tilde{\Psi}^{\bot}\rangle, e^{i\theta_2}|\tilde{\Phi}^{\bot}\rangle)
\label{app29}
\end{eqnarray}
is a smooth function on $\mathbb{S}_1\times \mathbb{S}_1$, where $\mathbb{S}_1$ denotes the complex circle on $\mathbb{C}^2$. By inserting $e^{i\theta_1}|\tilde{\Psi}^{\bot}\rangle$ and $e^{i\theta_2}|\tilde{\Phi}^{\bot}\rangle$ in Eq.(\ref{app29}), from the Fourier transformation and the linearity of $F_i$, it follows that
\begin{eqnarray}
\tilde{a}(e^{i\theta_1}|\tilde{\Psi}^{\bot}\rangle, e^{i\theta_2}|\tilde{\Phi}^{\bot}\rangle)
=
\tilde{a}(|\tilde{\Psi}^{\bot}\rangle,|\tilde{\Phi}^{\bot}\rangle).
\label{app30}
\end{eqnarray}
Moreover, we obtain
\begin{eqnarray}
&&\!\!\!\!\!\!\!\!(F\otimes I\otimes I)|\mu\rangle|\tilde{X}\rangle|\tilde{X}\rangle
=\tilde{a}(\frac{\alpha}{\sqrt{c_2}}+\frac{\beta}{\sqrt{c_2}})|\tilde{X}\rangle
|\phi\rangle|\psi\rangle,
\label{app31}
\\
&&\!\!\!\!\!\!\!\!(F\otimes I\otimes I)|\mu\rangle|\tilde{\Phi}^\bot\rangle|\tilde{\Psi}^{\bot}\rangle
=0,
\label{app32}
\\
&&\!\!\!\!\!\!\!\!(F\otimes I\otimes I)|\mu\rangle|\tilde{\Phi}^\bot\rangle|\tilde{X}\rangle
=\frac{\tilde{a}\alpha}{\sqrt{c_2}}|\tilde{\Phi}^\bot\rangle|\phi\rangle|\psi\rangle,
\label{app33}
\\
&&\!\!\!\!\!\!\!\!(F\otimes I\otimes I)|\mu\rangle|\tilde{X}\rangle|\tilde{\Psi}^\bot\rangle
=\frac{\tilde{a}\beta}{\sqrt{c_2}}|\tilde{\Psi}^\bot\rangle
|\phi\rangle|\psi\rangle.
\label{app34}
\end{eqnarray}
From Eqs.(\ref{app33}) and (\ref{app34}), $\tilde{a}$ should be independent of vectors $|\tilde{\Phi}^\bot\rangle$ and $|\tilde{\Psi}^\bot\rangle$. It means that $\tilde{a}$ is a constant. Similar proof may be followed for the conditions ${\rm tr}(\rho_X\rho_{\psi}\rho_{\psi}\rho_{\phi})=c_1>0$ and ${\rm tr}(\rho_X\rho_{\phi}\rho_{\phi}\rho_{\psi})=c_2>0$. This completes the proof.  $\hfill{} \Box$

The present proof also holds for $\rho_{\psi}$ and $\rho_{\phi}$ satisfying ${\rm tr}(\rho_X\rho_{\psi}\rho_{\phi}\rho_{\psi})=\lambda c_1$ and ${\rm tr}(\rho_X\rho_{\phi}\rho_{\psi}\rho_{\phi})=\lambda c_2$,
where $\lambda\in (0, \frac{1}{\max\{c_1,c_2\}}]$. The superpositions are then generated with a probability $p'=\lambda p$. From the uniqueness result, it is impossible to generate superpositions for all input states with a nonzero overlap with $\rho_{X}$.

We now present an explicit protocol to generate the superposition with a higher success probability \cite{10}. Let ${\cal G}(\rho)=G\rho G^\dag$, for a linear mapping ${\cal G}:\mathbb{H}^{4}\to \mathbb{H}$ defined by $G=G_2G_1$, where
\begin{eqnarray*}
G_1&=&\frac{\alpha}{\sqrt{c_1}}I\otimes I\otimes I\otimes I
+\frac{\beta}{\sqrt{c_2}}S_{1,3}\otimes S_{2,4},
\\
G_2&=&I\otimes \langle X|,
\end{eqnarray*}
$S_{i,j}$ denotes the swapping operation of the $i$-th and $j$-th state in $\mathbb{H}$ and $|X\rangle$ is the vector representative of $\rho_X$. The action of $G_2$ on the tensor of $|x_1\rangle|x_2\rangle|x_3\rangle|x_4\rangle$ is given
\begin{eqnarray}
I\otimes \langle X|x_1\rangle|x_2\rangle|x_3\rangle|x_4\rangle=|x_1\rangle(\langle X|x_2\rangle|x_3\rangle|x_4\rangle)
\label{app35}
\end{eqnarray}
for all $\rho_{x_i}\in \mathbb{H}$. With forward evaluations, we can obtain
\begin{eqnarray}
G(|\psi\rangle^{\otimes 2}|\phi\rangle^{\otimes 2})
=\alpha e^{i\theta_1}|\psi\rangle+
\beta e^{i\theta_2}|\phi\rangle
\label{app36}
\end{eqnarray}
which shows that $G(\rho_{\psi}\rho_{\psi}\rho_{\phi}\rho_{\phi}) G^\dag= \rho_{\varphi}$ up a global factor. ${\cal G}$ is trace non-increasing if and only if $GG^\dag\leq I\otimes I$, i.e,
\begin{eqnarray}
G^\dag G&=&
\frac{|\alpha|^2}{c_1}I\otimes \rho_X
+\frac{|\beta|^2}{c_2}\rho_X\otimes I
\nonumber
\\
&&
+\frac{\alpha\beta^*}{\sqrt{c_1c_2}}I\otimes \rho_X
(S_{1,3}\otimes S_{2,4})
\nonumber
\\
&&
+\frac{\alpha^*\beta}{\sqrt{c_1c_2}}(S_{1,3}\otimes S_{2,4})
(I\otimes \rho_X)
\label{app37}
\end{eqnarray}
The maximal eigenvalue of $G^\dag G$ is given by
\begin{eqnarray}
\lambda_{max}
&=&\max\{\frac{|\beta|^2}{c_2}+\frac{|\alpha|^2}{2c_1}
(\sqrt{\frac{4|\beta|^2}{c_2}+1}+1),
\nonumber
\\
&&|\frac{\alpha}{\sqrt{c_1}}+\frac{\beta}{\sqrt{c_2}}|^2\}.
\label{app38}
\end{eqnarray}
The largest $x\in \mathbb{R}^+$ satisfying $\tilde{\cal F}=x\cdot {\cal F}$ being non-increasing is $1/\lambda_{max}$. The success probability is
\begin{eqnarray}
p_{s}={\rm tr}(\tilde{\cal F}
(\rho_{\psi}\rho_{\psi}\rho_{\phi}\rho_{\phi}))
=
\frac{N^2_{\varphi}}{\lambda_{max}}.
\end{eqnarray}
Hence, $\tilde{p}_{s}\geq p_{s}$ if and only if
\begin{eqnarray}
\frac{1}{c_1}+\frac{1}{c_2}\geq \lambda_{max}
\end{eqnarray}
for $c_{i}\in(0,1]$ and $|\alpha|^2+|\beta|^2=1$.

\section*{Appendix G. Proof of Theorem 9}

Define $|v\rangle=\sqrt{c}(\sqrt{c_1}|0\rangle+\sqrt{c_2}|1\rangle)$, where $c$ is a normalization constant. Given two permutations $\tau_1, \tau_2\in \mathbb{P}_{2k-1}$, for all the states $\rho_{\Psi_{\tau_1(1,2,\cdots, 2k-1)}}$ satisfying ${\rm tr}(\rho_X\otimes\rho_{\Psi_{\tau_1(1,2,\cdots, 2k-1)}})=c_1>0$ and the states $\rho_{\Phi_{{\tau}_2(1,2,\cdots, 2k-1)}}$ satisfying ${\rm tr}(\rho_X\otimes \rho_{\Phi_{{\tau_1(1,2,\cdots, 2k-1)}}})=c_2>0$, we can define a CP map ${\cal F}_{\tau_1,\tau_2}$ such that
\begin{eqnarray}
{\cal F}_{\tau_1,\tau_2}(\rho_\mu\rho_{\psi}^{\otimes k} \rho_{\phi}^{\otimes k})=\rho_{\varphi},
\end{eqnarray}
where vector representative $|\varphi\rangle=\sqrt{r}(\alpha e^{i\theta_1}|\psi\rangle+
\beta e^{i\theta_2}|\phi\rangle)$ with $e^{i\theta_1}=\frac{\langle X|\Phi_{{\tau}_2(1,2,\cdots, 2k-1)}\rangle}{|\langle X|\Phi_{{\tau}_2(1,2,\cdots, 2k-1)}\rangle|}$ and $e^{i\theta_2}=\frac{\langle X|\Psi_{\tau_1(1,2,\cdots, 2k-1)}\rangle}{|\langle X|\Psi_{\tau_1(1,2,\cdots, 2k-1)}\rangle|}$ and $r$ is a normalized constant, and the vector $|\mu\rangle$ is given by $|\mu\rangle=\alpha|0\rangle+\beta|1\rangle$. Let
\begin{eqnarray}
{\cal F}_{\tau_1,\tau_2}={\cal F}_6\circ{\cal F}_5\circ {\cal F}_4\circ {\cal F}_3\circ {\cal F}_2\circ{\cal F}_1,
\end{eqnarray}
where ${\cal F}_j(\rho)=F_j\rho F^\dag_j, j=1,2,\cdots, 5$, ${\cal F}_6(\rho)={\rm tr}_{1,3,4,\cdots, nk+1}(\rho)$, and
\begin{eqnarray*}
F_1&=&|0\rangle\langle 0|\otimes I^{2k}+|1\rangle\langle 1|\otimes S_{2,k+2}\otimes I^{k-1},
 \\
F_2&=&|0\rangle\langle 0|\otimes I \otimes S_{\tau_1}+|1\rangle\langle 1| \otimes I^{2k},
\\
F_3&=&|0\rangle\langle 0|\otimes I^{2k}+|1\rangle\langle 1|\otimes I \otimes S_{\tau_2}
 \\
F_4&=&I_2\otimes I\otimes |X\rangle\langle X| ,
\\
F_5&=&|v\rangle\langle v|\otimes I^{nk},
\end{eqnarray*}
where $I^j$ denotes the identity operator on $\mathbb{H}^j (j=1,2, \cdots, nk)$, $I_2$ denotes the identity operator on $\mathbb{C}^2$, $S_{\tau_i}$ is a swapping operator induced by the permutation $\tau_i$ in $\mathbb{P}_{2k-1}$ and performed on the last $2k-1$ subsystems, $i=1, 2$. It is easy to get the result by forward evaluations from the input states $\rho_{v}\otimes \rho_{\psi}^{\otimes k}\otimes \rho_{\phi}^{\otimes k}$.

\section*{Appendix H: Proof of Corollary 5}

Define an ancillary vector
\begin{eqnarray}
|v\rangle=\sqrt{c}(\sum_{j=0}^{n-1}\frac{\sqrt{c_j}}
{\prod_{j=0}^{n-1}\sqrt{c_j}}|j\rangle),
\end{eqnarray}
where $c$ is a normalization constant. Let
\begin{eqnarray}
{\cal F}={\cal F}_5\circ {\cal F}_4\circ {\cal F}_3\circ {\cal F}_2\circ{\cal F}_1,
\end{eqnarray}
where ${\cal F}_j(\rho)=F_j\rho F^\dag_j, j=1,2,\cdots, 4$, ${\cal F}_5(\rho)={\rm tr}_{1,3,4,\cdots, nk,nk+1}(\rho)$, and
\begin{eqnarray*}
F_1&=&|0\rangle\langle 0|\otimes I^{nk}+\sum_{j=1}^{n-1}|j\rangle\langle j|\otimes S_{2,jk+2},
 \\
F_2&=&\sum_{j=0}^{n-1}|j\rangle\langle j|\otimes S_{\tau_j},
\\
F_3&=&I_n\otimes I\otimes |X\rangle\langle X|,
\\
F_4&=&|v\rangle\langle v|\otimes I^{nk},
\end{eqnarray*}
where $I^j$ denotes the identity operator on Hilbert space $\mathbb{H}^j (j=1,2, \cdots, nk)$, $I_n$ denotes the identity operator on $\mathbb{C}^2$, $S_{\tau_j}$ is a swapping operator induced by the permutation $\tau_j$ in $\mathbb{P}_{nk-1}$ and performed on the last $nk-1$ subsystems, $j=0, 1,  \cdots, n-1$. $F_1$ is used to change the vector $|v\rangle (\otimes_{i=1}^n |\psi_i\rangle^{\otimes k})$ into $\sqrt{c}\sum_{j=0}^{n-1}\sqrt{c_j}|j\rangle|\psi_j\rangle|\Psi^{j}_{1,2,\cdots, nk-1}\rangle$. $F_2$ is used to change the vector $\sqrt{c}\sum_{j=0}^{n-1}\sqrt{c_j}|j\rangle|\psi_j\rangle|\Psi^{j}_{1,2,\cdots, nk-1}\rangle$ into $\sum_{j=0}^{n-1}\sqrt{c_j}|j\rangle|\psi_j\rangle|\Psi^{j}_{\tau_j(1,2,\cdots, nk-1)}\rangle$, where $|\Psi^{j}_{1,2,\cdots, nk-1}\rangle=\otimes_{t=1}^n|\psi_t\rangle^{\otimes k_{jt}}$ with integers $k_{jt}$ satisfying $k_{jt}=k$ for $j\not=t$ and $k_{jt}=k-1$ for $j=t$.
From forward evaluations, we can prove the results using the followed measurements induced by the operators $|X\rangle\langle X|$ and $|v\rangle\langle v|$ \cite{10}.


\begin{thebibliography}{99}
\bibitem{Form}
For linearly independent vectors $|\psi_1\rangle, |\psi_1\rangle, \cdots, |\psi_m\rangle$, there exists a vector $|\tilde{\psi}\rangle$ such that $\langle \psi_i|\tilde{\psi}\rangle=c$, $i=1, 2, \cdots, m$.
\bibitem{56} It can be equvilently obtained from ${\rm tr}({\cal U}(\rho_{\Omega_i}){\cal U}(\rho_{\Omega_j})={\rm tr}(\rho_{\psi_i}^{\otimes k}\rho_{\psi_j}^{\otimes k})=|a_{ij}|^{2k}$, where $\rho_{\Omega_{i}}=\rho_{\psi_i}^{\otimes k}\rho_{\Sigma}\rho_{P_0}$, $\rho_{\Omega_{j}}=\rho_{\psi_j}^{\otimes k}\rho_{\Sigma}\rho_{P_0}$ and $a_{ij}=\langle \psi_i|\psi_j\rangle$.
\end{thebibliography}
\end{document}